\documentclass[onecolumn,12pt]{article}
\usepackage{graphicx}
\usepackage{color}
\newcommand\simlt{\hspace{0.3em}\raisebox{0.4ex}{$<$}\hspace{-0.75em}\raisebox{-.7ex}{$\sim$}\hspace{0.3em}} 
\newcommand\simgt{\hspace{0.3em}\raisebox{0.4ex}{$>$}\hspace{-0.75em}\raisebox{-.7ex}{$\sim$}\hspace{0.3em}} 

\newcommand{\beginsupplement}{%
        \setcounter{table}{0}
        \renewcommand{\thetable}{S\arabic{table}}%
        \setcounter{figure}{0}
        \renewcommand{\thefigure}{S\arabic{figure}}%
     }

\begin{document}
\baselineskip 14pt

\title{Distortion of Magnetic Fields in the Dense Core CB81 (L1774, Pipe 42) in the Pipe Nebula}
\date{Ver.6}
\author{Ryo Kandori$^{1}$, Motohide Tamura$^{1,2,3}$, Masao Saito$^{3}$, Kohji Tomisaka$^{3}$, \\
Tomoaki Matsumoto$^4$, Ryo Tazaki$^{5}$, Tetsuya Nagata$^{6}$, Nobuhiko Kusakabe$^{1}$, \\
Yasushi Nakajima$^{7}$, Jungmi Kwon$^{8}$, Takahiro Nagayama$^{9}$, and Ken'ichi Tatematsu$^{3}$\\
{\small 1. Astrobiology Center of NINS, 2-21-1, Osawa, Mitaka, Tokyo 181-8588, Japan}\\
{\small 2. Department of Astronomy, The University of Tokyo, 7-3-1, Hongo, Bunkyo-ku, Tokyo, 113-0033, Japan}\\
{\small 3. National Astronomical Observatory of Japan, 2-21-1 Osawa, Mitaka, Tokyo 181-8588, Japan}\\
{\small 4. Faculty of Sustainability Studies, Hosei University, Fujimi, Chiyoda-ku, Tokyo 102-8160}\\
{\small 5. Astronomical Institute, Graduate School of Science Tohoku University,}\\
{\small 6-3 Aramaki, Aoba-ku, Sendai 980-8578, Japan}\\
{\small 6. Kyoto University, Kitashirakawa-Oiwake-cho, Sakyo-ku, Kyoto 606-8502, Japan}\\
{\small 7. Hitotsubashi University, 2-1 Naka, Kunitachi, Tokyo 186-8601, Japan}\\
{\small 8. Institute of Space and Astronautical Science, Japan Aerospace Exploration Agency,}\\
{\small 3-1-1 Yoshinodai, Chuo-ku, Sagamihara, Kanagawa 252-5210, Japan}\\
{\small 9. Kagoshima University, 1-21-35 Korimoto, Kagoshima 890-0065, Japan}\\
{\small e-mail: r.kandori@nao.ac.jp}}
\maketitle

\begin{abstract}
The detailed magnetic field structure of the starless dense core CB81 (L1774, Pipe 42) in the Pipe Nebula was determined based on near-infrared polarimetric observations of background stars to measure dichroically polarized light produced by magnetically aligned dust grains in the core. The magnetic fields pervading CB81 were mapped using 147 stars and axisymmetrically distorted hourglass-like fields were identified. On the basis of simple 2D and 3D magnetic field modeling, the magnetic inclination angles in the plane-of-sky and line-of-sight directions were determined to be $4^{\circ} \pm 8^{\circ}$ and $20^{\circ} \pm 20^{\circ}$, respectively. The total magnetic field strength of CB81 was found to be $7.2 \pm 2.3$ $\mu{\rm G}$. Taking into account the effects of thermal/turbulent pressure and magnetic fields, the critical mass of CB81 was calculated to be $M_{\rm cr}=4.03 \pm 0.40$ M$_{\odot}$, which is close to the observed core mass of $M_{\rm core}=3.37 \pm 0.51$ M$_{\odot}$. We thus conclude that CB81 is in a condition close to the critical state. In addition, a spatial offset of $92''$ was found between the center of magnetic field geometry and the dust extinction distribution; this offset structure could not have been produced by self-gravity. The data also indicate a linear relationship between polarization and extinction up to $A_V \sim 30$ mag going toward the core center. This result confirms that near-infrared polarization can accurately trace the overall magnetic field structure of the core.
\end{abstract}

\vspace*{0.3 cm}

\clearpage

\section{Introduction}
CB81 (catalogued by Clemens \& Barvainis 1988, also referred to as Pipe 42 by Alves et al. 2007, and as L1774 in the catalog of Lynds 1962) is a dense core in the Pipe Nebula, representing a nearby dark cloud complex having a filamentary shape located at a distance of $d=130^{+24}_{-58}$ pc (Lombardi et al. 2006, see also Alves \& Franco 2007; Dzib et al. 2018). A search for young stellar objects (YSOs) was previously conducted using the {\it Spitzer} telescope at 24 $\mu$m and 70 $\mu$m, covering the entire surface of the Pipe Nebula, and no YSOs were reported toward CB81 (Forbrich et al. 2009). A search for (diskless) pre-main-sequence (PMS) stars based on X-ray observations also resulted in no such objects being detected in the region of CB81 (Forbrich et al. 2010). Thus, CB81 can be considered to be a starless dense core, despite a report of the possible detection of a CO outflow (Yun \& Clemens 1992, 1994) near an IRAS source toward CB81. NH$_3$ observations were conducted toward CB81 in radio wavelengths (Rathborne et al. 2008). The temperature of the core was reported to be $T_{\rm kin} = 12.4 \pm 0.4$ K. The velocity dispersions, $\sigma_v$, associated with the NH$_3$ (1,1) and NH$_3$ (2,2) lines were determined to be 0.11 km s$^{-1}$ and 0.09 km s$^{-1}$, respectively. Using a value of 0.1 km s$^{-1}$ for the NH$_3$ velocity dispersion for the core, the calculated turbulent velocity dispersion, $\sigma_{\rm turb}$, is 0.0627 km s$^{-1}$. 
Franco et al. (2010) employed $R$ band linear polarimetry toward the CB81 region in the Pipe Nebula, and reported a polarization angle of $\theta_{R} = 158.1^{\circ} \pm 5.35^{\circ}$. The magnetic field direction in the CB81 region is roughly perpendicular to the global filament of the Pipe Nebula. 
\par
In addition to the studies described above, many surveys and statistical analyses have been performed toward the Pipe Nebula, including extinction map studies (Lombardi et al. 2006; Alves et al. 2007; Lada et al. 2008; Rom\'{a}n-Z\'{u}\~{n}iga et al. 2010), dust emission studies using data acquired from the {\it Herschel} satellite (Peretto et al. 2012), radio wave molecular line studies (Onishi et al. 1999; Muench et al. 2007; Rathborne et al. 2009), polarimetric studies (Alves et al. 2008), and dust temperature studies (Forbrich et al. 2014; Hasenberger et al. 2018). However, there have only been a few research projects focusing specifically on CB81. CB81 is located in the \lq \lq stem'' region of the Pipe Nebula, and appears bright in far infrared maps generated by the {\it Herschel} satellite. This core is also evident in a previously reported dust extinction map (Rom\'{a}n-Z\'{u}\~{n}iga et al. 2010), with a peak $A_V$ of greater than 20 mag. CB81 is the densest core in the \lq \lq stem'' region and is relatively isolated from neighboring cores (see Figure 4 of Rom\'{a}n-Z\'{u}\~{n}iga et al. 2010). Moreover, the Pipe Nebula has a rich stellar backdrop consisting of galactic bulge stars. These characteristics make CB81 an ideal object with regard to studying the magnetic fields of dense cores. 
\par
Since the search of the hourglass-shaped magnetic fields toward dense cores is rather limited (Kandori et al. 2017a, hereafter Paper I; Kandori et al. 2019a), we chose CB81 for the target in our near-infrared (NIR) polarimetric surveys of dense cores. To increase the number of cores with hourglass-shaped magnetic fields is important to take statistics of the total magnetic field strength, mass-to-flux ratio, and polarization--extinction relationship through the three dimensional (3D) modeling of the hourglass fields (Kandori et al. 2017b, hereafter Paper II; Kandori et al. 2017c, hereafter Paper III; Kandori et al. 2018b; Kandori et al. 2018c). These statistics are closely related to the study of the initial conditions of star formation, the core formation process, and the magnetically aligned dust in dense environments. 
\par
In the present study, wide-field background star polarimetry at NIR wavelengths was conducted for CB81 and the Stokes I data in the $H$ and $K_s$ bands were used to map the dust extinction characteristics of CB81 as a means of determining the density structure of the core (Appendix). The plane-of-sky magnetic field structure was ascertained using several thousands of stars in and around the core radius, while removing contamination from the ambient field component. In addition, the total magnetic field strength of the core was estimated based on the Davis-Chandrasekhar-Fermi method (Davis 1951; Chandrasekhar \& Fermi 1953) and the 3D magnetic field modeling of the core. Using the resulting magnetic field information, the kinematical stability of CB81 is discussed herein. The relationship between polarization and extinction ($P$--$A$) for CB81 was derived by subtracting ambient polarization and correcting for the depolarization effect caused by distortions of the magnetic field shape and the line-of-sight magnetic inclination angle. The resulting linear $P$--$A$ relationship indicates that the observed polarization reflects the overall magnetic field structure in the core. 
\section{Observations and Data Reduction}
We observed CB81 using the $JHK_s$-simultaneous imaging camera SIRIUS (Nagayama et al. 2003) with the associated polarimetry mode SIRPOL (Kandori et al. 2006) on the IRSF 1.4-m telescope at the South African Astronomical Observatory (SAAO). IRSF/SIRPOL is a useful instrument for NIR polarization surveys because it simultaneously provides wide-field $JHK_s$ polarization images ($7.\hspace{-3pt}'7 \times 7.\hspace{-3pt}'7$ with a scale of 0$.\hspace{-3pt}''$45 ${\rm pixel}^{-1}$). 
The uncertainty in the polarization degree values resulting from sky variations during exposures is typically $0.3\%$. The uncertainty in the determination of the polarization angle origin (i.e., the correction angle) is less than 3$^{\circ}$ (Kandori et al. 2006; Kusune et al. 2015). 
\par
We observed the polarized standard star RCrA\#88 ($P_H = 2.73\% \pm 0.07\%$, $\theta_H = 92^{\circ} \pm 1^{\circ}$, Whittet et al. 1992) on July 13, 2017, during the two months run of our observations. Values of $P_H = 2.82\% \pm 0.09\%$ and $\theta_H = 91.9^{\circ} \pm 0.9^{\circ}$ were obtained, which are consistent with data in the literature. 
\par
Observations of CB81 were conducted on the nights of June 15 and 24, 2017. A number of 15s exposures were performed at four half-waveplate angles (in the sequence of $0^{\circ}$, $45^{\circ}$, $22.5^{\circ}$, and $67.5^{\circ}$) at ten dithered positions (equivalent to one set). The total integration time was 3000s (20 sets) per waveplate angle. 
\par
The data thus acquired were reduced using the Interactive Data Language (IDL) software. This reduction included flat-field correction with twilight flat frames, median sky subtraction and frame combining after registration and was performed in the same manner as described by Kandori et al. (2007). 
Note that the accuracy of the calibration based on twilight flat images is discussed in the Appendix. 
Point sources having peak intensities greater than $10 \sigma $ above the local sky background could be detected on the Stokes $I$ images, and aperture polarimetry was performed for a number of sources for each waveplate position angle image ($I_{0}$, $I_{45}$, $I_{22.5}$, and $I_{67.5}$). In this manner, 5159, 9576, and 5895 sources were detected in the $J$, $H$, and $K_s$ bands, respectively. The aperture radius equaled the full width at half maximum (FWHM) of the stars (2.8 pixels for the $J$, $H$, and $K_s$ bands), and the sky annulus was set to 10 pixels with a 5 pixel width. The limiting magnitudes were 18.5, 18.0, and 17.0 mag in the $J$, $H$, and $K_s$ bands, respectively. A relatively small aperture radius was used to suppress flux contamination from neighboring stars. We did not use psf-fitting photometry because the goodness of fit for each star on different waveplate angle images can cause systematic errors in polarization measurements. 
%
Sources with photometric errors greater than 0.1 mag were rejected. 
\par
The Stokes parameter for each star was derived using the relationships $I = (I_{0} + I_{45} + I_{22.5} + I_{67.5})/2$, $Q = I_{0} - I_{45}$, and $U = I_{22.5} - I_{67.5}$. The polarization degree, $P$, and polarization angle, $\theta$, were determined from the equation $P = \sqrt{Q^2 + U^2}/I$ and $\theta = 0.5 {\rm atan}(U/Q)$. Because $P$ is a positive quantity, the resulting $P$ values tend to be overestimated, especially in the case of low $S/N$ sources. We corrected for this bias using the equation $P_{\rm db} = \sqrt{P^2 - \delta P^2}$ (Wardle \& Kronberg 1974). 
In the present study, we discuss the results obtained working in the $H$ band, for which dust extinction effects are less severe than in the $J$ band, and polarization efficiency is greater than in the $K_s$ band. 

\section{Results and Discussion}
\subsection{Distortion of Magnetic Fields}
Figure 1 presents the resulting polarization vector map for CB81 in the $H$ band. Here, stars for which $P_H / \delta P_H \ge 5$ are shown. CB81 is located slightly to the east from the center of the image, appearing as a region of dark obscuration. The white circle indicates the core radius ($R=177''$) determined from NIR extinction studies (see Appendix). 

Although the polarization distribution is relatively uniform outside the core radius, a kink or bent structure is evident in the polarization vectors inside the radius. The polarization vectors located outside the core are regarded as off-core polarizations generated by the ambient medium. As shown in Figures 2 and 3, the off-core vectors were found to have values of $P_{H,{\rm off}} = 2.40 \pm 0.60$\% and $\theta_{H,{\rm off}} = 152.9^{\circ} \pm 8.9^{\circ}$ (representing the median value and one standard deviation). The off-core polarizations are relatively well ordered (Figure 3) and have similar polarization degrees (Figure 2). 
The polarization angle, $\theta_{H, {\rm off}}$, is consistent with the value obtained based on observations in the $R$ band toward the CB81 region, $\theta_{R} = 158.1^{\circ} \pm 5.35^{\circ}$ (Franco et al. 2010). 
Following the same procedure described in our previous paper (Paper I), we fitted the off-core vectors on the sky plane in $Q/I$ and $U/I$. The distributions of the $Q/I$ and $U/I$ values were subsequently modeled as $f(x,y)=A+Bx+Cy$, where $x$ and $y$ are the pixel coordinates, and $A$, $B$, and $C$ are the fitting parameters. 
Note that we did not employ higher order spatial fitting (e.g., second order fitting). The F-test based on residuals of the first and second order fittings resulted in significance values of $58.6\%$ for $Q/I$ and $66.3\%$ for $U/I$, both of which greatly exceed the 5\% or 1\% level, and these fittings cannot be distinguished. Comparing the plane fitting to the median value obtained from subtraction analysis (i.e., the zero order), the F-test gives significance values of $0.6\%$ for $Q/I$ and $9.3 \times 10^{-8}\%$ for $U/I$. Therefore, it is evident that the plane fitting process is suitable. 
\par
The off-core vectors generated in this work are provided in Figure 4. The regression vectors of the off-core polarization components were subsequently subtracted from the original vectors in order to isolate the polarization vectors associated with CB81 (Figure 5). Comparing Figures 1 and 5, the ordered fields going from the north-west to south-east disappear and a distorted polarization distribution appears in and around the core. Following the subtraction, the polarization degree of the off-core vectors is successfully suppressed to give a value close to 0\% (Figure 2), while the polarization angle becomes randomly distributed (Figure 3). Therefore, the subtraction analysis was evidently successful. 
\par
The distance to the off-core stars was examined using the Gaia DR2 data set (Gaia Collaboration et al. 2016, 2018) and all the Gaia DR2 sources (with faint limiting magnitudes of $G \sim 20$ mag) distributed within 4 arcmin from CB81 were found to be located in the background to the core. Thus, when determining the ambient off-core polarization vectors, contamination from foreground stars could be ignored. 
\par
Interestingly, in Figure 5, the distribution of the axisymmetrically distorted fields seems to have been offset based on the dust obscuration distribution. We thus treated the center of magnetic field geometry as a free parameter in the model fitting process presented in the subsequent Parabolic Model section. 
\par
It is notable that our analysis of the subtraction of ambient off-core polarization component may look wrong, because the analysis may remove the very low polarization angle dispersion (i.e., strong magnetic field component that laces both the surrounding diffuse interstellar medium and dense molecular cloud). However, this is not the case. The Stokes parameters observed toward the core can be written as $(Q/I)_{\rm obs} = q_{\rm obs} = q_{\rm ISM} + q_{\rm core}$ and $(U/I)_{\rm obs} = u_{\rm obs} = u_{\rm ISM} + u_{\rm core}$, where $q_{\rm ISM}$ and $u_{\rm ISM}$ show the polarization component for the diffuse surrounding medium and $q_{\rm core}$ and $u_{\rm core}$ show the polarization component arisen in the core. The ambient subtraction analysis just isolates $q_{\rm core}$ and $u_{\rm core}$, and this analysis does not affect the alignment of polarization vectors solely associated with the core. 

\subsection{Parabolic Model}
The most probable configuration for the magnetic field lines, estimated using a parabolic function and its shift and rotation, is shown in Figure 6 (solid white lines). In this figure, 147 polarization vectors having $P_H \ge 0.5$ \% are included and the field of view is the same as the diameter of CB81 ($354''$) in the $\delta$ direction, with a value of $294''$ in the $\alpha$ direction. The size of the field of view therefore different in the $\alpha$ and $\delta$ directions. This is the case because the center of the distorted field is located slightly to the east relative to the center of obscuration, and the eastern end of the distorted field was not fully incorporated into field of view during the observations. A comparison of the distorted field to the $A_V$ distribution is presented in Figure 7, in which the white and red crosses indicate the center of the distorted field and the centroid center of the $A_V$ distribution, respectively (see Appendix). It is evident from this image that there is a displacement between the center of the magnetic field structure and the center of mass distribution. 
\par
The best-fit parameters were determined to be $\theta_{\rm mag}=4^{\circ} \pm 8^{\circ}$ and $C=1.43(\pm 0.68) \times 10^{-5}$ pixel$^{-2}$ ($=7.06 \times 10^{-5}$ arcsec$^{-2}$ $=6.4 \times 10^{-9}$ AU$^{-2}$) for the parabolic function $y=g+gCx^2$, where $g$ specifies magnetic field lines, $\theta_{\rm mag}$ is the position angle of the magnetic field direction (from north through east), and $C$ determines the degree of curvature of the function. 
Note that we used the parabolic function in the $90^{\circ}$-rotated form so that, in the case that $\theta_{\rm mag}$ is $0^{\circ}$, the $x$-direction corresponds to the direction of declination. 
\par
A parabolic function was employed because this is the simplest means of approximating the analytically described hourglass-shaped magnetic field model (Mestel 1966; Ewertowski \& Basu 2013; Myers et al. 2018; see also Kandori et al. 2019b, hereafter Paper VI, for a comparison of the parabolic function to the hourglass model). In a subsequent paper, we intend to use the Myers model to fit the same data and to discuss the core formation mechanism. 
\par
The observational error associated with each star was taken into account when calculating $\chi^2 = (\sum_{i=1}^n (\theta_{\rm obs,{\it i}} - \theta_{\rm model}(x_i,y_i))^2 / \delta \theta_i^2$, where $n$ is the number of stars, $x$ and $y$ are the coordinates of these stars, $\theta_{\rm obs}$ and $\theta_{\rm model}$ denote the polarization angles from observations and from the model, and $\delta \theta_i$ is the observational error) during the fitting procedure. The coordinate origin of the parabolic function is R.A.=17$^{\rm h}$22$^{\rm m}$47$.\hspace{-3pt}^{\rm s}$86, Decl.=-27$^{\circ}$04$'$40$.\hspace{-3pt}''5$ (J2000), as determined by searching for the minimum $\chi^2$ for the variables $x$ and $y$ in and around the core. The center coordinate of the core (that is, the centroid center of the $A_V$ distribution, see Appendix for the derivation of the $A_V$ map) is R.A.=17$^{\rm h}$22$^{\rm m}$41$.\hspace{-3pt}^{\rm s}$42, Decl.=-27$^{\circ}$05$'$12$.\hspace{-3pt}''8$ (J2000). The angular distance between these two centers is $92''$, which corresponds to roughly half the core radius. 
The uncertainties to determine the center coordinates were estimated to be $18.0''$ and $16.7''$ for the magnetic center and the extinction center, respectively. For the latter value, we used the difference of the centroid center between the $A_V$ distribution and the {\it Herschel} $500$ $\mu$m emission map (Peretto et al. 2012; Roy et al. 2019; Andr\'{e} et al. 2010) as the 1-sigma uncertainty. 
We discuss the origin of this displacement in Section 3.4. 
\par 
The parabolic fitting was evidently satisfactory, since the standard deviation of the residual angles, $\theta_{\rm res} = \theta_{\rm obs} - \theta_{\rm fit}$, was smaller when using the parabolic function ($\delta \theta_{\rm res}=20.03^{\circ} \pm 0.78^{\circ}$, see Figure 8) than in the case of a uniform field ($\delta \theta_{\rm uni}=30.78^{\circ} \pm 0.69^{\circ}$). 
The uncertainties associated with $\delta \theta_{\rm res}$ and $\delta \theta_{\rm uni}$ were estimated using the bootstrap method. In this process, a random number following a normal distribution with the same width as the observational error was added for each star, and fitting was performed using the parabolic function. This process was repeated 1000 times to ascertain the dispersion of the resulting $\delta \theta_{\rm res}$ distribution. 
Comparing $\delta \theta_{\rm res}$ and $\delta \theta_{\rm uni}$, there is a statistically significant difference, suggesting that a distorted field is most likely present. This was also confirmed from an F-test with a significance of $3.3 \times 10^{-7}\%$, indicating that $\delta \theta_{\rm res}$ and $\delta \theta_{\rm uni}$ are statistically different. The intrinsic dispersion, $\delta \theta_{\rm int} = (\delta \theta_{\rm res}^2 - \delta \theta_{\rm err}^2)^{1/2}$, estimated using the parabolic fitting was found to be $18.65^{\circ} \pm 0.84^{\circ}$ (0.3260 radian), where $\delta \theta_{\rm err}$ is the standard deviation of the observational error in the polarization measurements. 
\par
CB81 is the third starless dense core associated with axisymmetrically distorted hourglass-like magnetic fields. 
The resulting magnetic curvature of $C=6.4 \times 10^{-9}$ AU$^{-2}$ is similar to those reported for FeSt 1-457 ($C=3.0 \times 10^{-9}$ AU$^{-2}$, Paper I) and B68 ($C=7.0 \times 10^{-9}$ AU$^{-2}$, Kandori et al. 2019), suggesting that similar mechanisms were responsible for the hourglass-like distortions in those dense cores/globules. 
\par
Assuming frozen-in magnetic fields, the intrinsic dispersion of the magnetic field direction, $\delta \theta_{\rm int}$, can be attributed to perturbation of the Alfv\'{e}n wave by turbulence. The strength of the plane-of-sky magnetic field (${B}_{\rm pos}$) can be estimated from the relationship ${B}_{\rm pos} = {C}_{\rm corr} (4 \pi \rho)^{1/2} \sigma_{\rm turb} / \delta \theta_{\rm int}$, where $\rho$ and $\sigma_{\rm turb}$ are the mean density of the core and the turbulent velocity dispersion (Davis 1951; Chandrasekhar \& Fermi, 1953), and ${C}_{\rm corr} = 0.5$ is a correction factor suggested by theoretical studies (Ostriker et al., 2001, see also, Padoan et al. 2001; Heitsch et al. 2001; Heitsch 2005; Matsumoto et al. 2006). 
Using the mean density $(\rho = 3.93 (\pm 0.79) \times 10^{-20}$ g cm$^{-3}$, see Appendix) and turbulent velocity dispersion ($\sigma_{\rm turb} = 0.0627$ km s${}^{-1}$) from Rathborne et al. (2008) and the $\delta \theta_{\rm int}$ derived in this study, a weak magnetic field is obtained as the lower limit of the total field strength ($|B|$): $B_{\rm pos} = 6.8$ $\mu {\rm G}$. 
For the purpose of error estimation, we assumed that the turbulent velocity dispersion value was accurate within 30\%. Thus, the uncertainty in $B_{\rm pos}$ was estimated to be 2.2 $\mu$G. 

\subsection{3D Magnetic Field}
The 3D magnetic field modeling was performed following the procedure described in a previous paper (Paper II, see also Paper VI). 
The 3D version of the simple parabolic function, $z(r, \varphi, g) = g + gC{r}^{2}$ in the cylindrical coordinate system $(r, z, \varphi)$ was used for the modeling of the core magnetic fields, where $g$ specifies the magnetic field lines, $C$ is the curvature of the lines, and $\varphi$ is the azimuth angle (measured in the plane perpendicular to $r$). Using this function, the magnetic field lines are axially symmetric around the $r$ axis, and the value of $z(r, \varphi, g)$ is not affected by the parameter $\varphi$. 
\par
The 3D model was virtually observed after rotating in the line-of-sight ($\gamma_{\rm mag}$) and the plane-of-sky ($\theta_{\rm mag}$) directions. The analysis was performed in the same manner described in Section 3.1 of Paper VI (see also Sections 2 and 3.1 of Paper II). The resulting polarization vector maps for the 3D parabolic model for various $\gamma_{\rm mag}$ are shown in Figure S1 of Paper VI. The density distribution of the model core was calculated based on the Bonnor--Ebert sphere model (Ebert 1955; Bonnor 1956) with a solution parameter of $\xi_{\rm max} = 13.0$ (see Appendix). The density distribution of CB81 was shifted from the center of the magnetic field configuration, as described in Sections 3.1 and 3.2, and so the Bonnor--Ebert density distribution in the 3D model was shifted by $92''$ to be consistent with observations. 
The model map can be compared with observations for each parameter set (see Figure S1 of Paper VI). Since the polarization distributions in the model core differ from one another depending on the viewing angle toward the line-of-sight, $\chi^{2}$ fitting of these distributions with the observational data can be used to restrict the line-of-sight magnetic inclination angle and 3D magnetic curvature. 
\par
Figure 9 summarizes the distribution of $\chi^2_\theta = ( \sum_{i=1}^n (\theta_{\rm obs,{\it i}} - \theta_{\rm model}(x_i,y_i))^2 / \delta \theta_i^2$, where $n$ is the number of stars, $x$ and $y$ show the coordinates of stars, $\theta_{\rm obs}$ and $\theta_{\rm model}$ denote the polarization angle from observations and the model, and $\delta \theta_i$ is the observational error) calculated by using the model and observed polarization angles. The optimal magnetic curvature parameter, $C$, was determined at each inclination angle $\gamma_{\rm mag}$ to obtain $\chi^2_\theta$. 
\par
From the polarization angle fitting, it is clear that the large inclination angle ($\gamma_{\rm mag} \simgt 30^{\circ}-40^{\circ}$) is unlikely. Note that the large $\gamma_{\rm mag}$ model shows a radial polarization pattern that does not match observations. The distribution of $\chi^2_\theta$ is relatively flat for the region $\gamma_{\rm mag} \simlt 20^{\circ}$. The minimization point for $\chi^2_\theta$ is $\gamma_{\rm mag} = 20^{\circ}$ with an uncertainty of $24^{\circ}$. Since $\chi^2_\theta$ increases rapidly from $\gamma_{\rm mag} = 20^{\circ}$ on going toward larger inclination angles, the uncertainty in this direction should be less than $24^{\circ}$. We can therefore conclude that the $\gamma_{\rm mag}$ value is $20^{\circ}$ with an uncertainty of $20^{\circ}$. 
The magnetic curvature obtained at $\gamma_{\rm mag} = 20^{\circ}$ is $C = 2.5 \times 10^{-4}$ ${\rm arcsec}^{-2}$. Note that the $\chi^2$ values in Figure 9 are relatively large. This result can likely be attributed to scatter of the polarization angles caused by Alfv\'{e}n waves, which cannot be included in the observational error term in the calculation of $\chi^2$. 
%
\par
Figure 10 shows the best-fit 3D parabolic model along with the observed polarization vectors. The direction of the model vectors generally agrees with the observations. The standard deviation of the differences in the plane-of-sky polarization angles between the 3D model and the observations is $15.24^{\circ}$, which is less than the 2D fitting result as well as that for a uniform field ($30.78^{\circ}$). 
\par
Figure 11 presents the same observational data as Figure 10 but with the background image processed using the line integral convolution (LIC) technique (Cabral \& Leedom 1993). Here, we used the publicly available IDL code developed by Diego Falceta-Gon\c{c}alves. The direction of the LIC \lq \lq texture'' is parallel to the magnetic field direction, and the background image is based on the polarization degree of the model core.

\subsection{Magnetic Properties of the Core}
Using the determined magnetic inclination angle, $\gamma_{\rm mag}$, of $20^{\circ}$, the total magnetic field strength of CB81 was found to be $B_{\rm pos}/\cos (\gamma_{\rm mag})=6.8/\cos(20^{\circ})=7.2 \pm 2.3$ $\mu{\rm G}$. 
The magnetic support of the core against gravity can be investigated using the parameter ${\lambda} = ({M}/{\Phi})_{\rm obs} / ({M}/{\Phi})_{\rm critical}$, which represents the ratio of the observed mass-to-magnetic flux ratio to a critical value, $(2\pi {\rm G}^{1/2})^{-1}$, suggested by theory (Mestel \& Spitzer 1956; Nakano \& Nakamura 1978). 
We determined a value of $\lambda = 4.04 \pm 0.80$ (that is, a magnetically supercritical value). The magnetic critical mass of the core of $0.83 \pm 0.25$ ${\rm M}_{\odot}$ was far lower than the observed core mass of $M_{\rm core}=3.37 \pm 0.51$ ${\rm M}_{\odot}$. 
%
\par
Although CB81 was found to be magnetically supercritical, this does not necessarily indicate that the core was in an unstable state. 
The critical mass of CB81, taking into account both magnetic and thermal/turbulent support effects equal to $M_{\rm cr} \simeq M_{\rm mag}+M_{\rm BE}$ (Mouschovias \& Spitzer 1976; Tomisaka, Ikeuchi, \& Nakamura 1988; McKee 1989), was determined to be $0.83+3.20=4.03 \pm 0.40$ ${\rm M}_{\odot}$, where $3.20 \pm 0.32$ ${\rm M}_{\odot}$ is the Bonnor--Ebert mass calculated using a kinematic temperature value of 12.4 K for the core (Rathborne et al. 2008), a turbulent velocity dispersion of $0.067$ km ${s}^{-1}$ (calculated from Rathborne et al. 2008), and an external pressure of $4.33 (\pm 0.75) \times 10^4$ K cm$^{-3}$ (see Appendix). 
Therefore, we conclude that CB81 is in a condition not far from the critical state. This critical (or marginally stable) state of CB81 is consistent with the starless characteristics of the core. 
\par
\par 
The relative importance of magnetic fields with regard to the support of the core was investigated using the ratios of the thermal and turbulent energy to the magnetic energy, ${\beta} \equiv 3{C}_{\rm s}^{2}/{V}_{\rm A}^{2}$ and ${\beta}_{\rm turb} \equiv {\sigma}_{\rm turb,3D}^{2}/{V}_{\rm A}^{2} = 3{\sigma}_{\rm turb,1D}^{2}/{V}_{\rm A}^{2}$, where ${C}_{\rm s}$, ${\sigma}_{\rm turb}$, and ${V}_{\rm A}$ denote the isothermal sound speed at 12.4 K, the turbulent velocity dispersion, and the Alfven velocity. 
These ratios were found to be ${\beta} \sim 12.4$ and ${\beta}_{\rm turb} \sim 1.12$. 
Thermal support is therefore dominant in CB81, with a small contribution from static magnetic fields and turbulence. The magnetic field is weak and the turbulence appears to have largely dissipated. 
\par
The $A_V$ distribution (Figure S1, see Appendix) of CB81 is almost circular from the center to boundary, although there is a filamentary tail-like structure extending toward the north-west. The plane-of-sky magnetic axis (PA=$4^{\circ}$) of CB81 is roughly perpendicular to this filamentary structure, and the circular shape of CB81 is consistent with its weak magnetic field strength. While a magneto-hydrostatic configuration can produce a flattened inner density distribution whose major axis is perpendicular to the direction of magnetic fields (Tomisaka, Ikeuchi, \& Nakamura 1988), the resulting density distribution can be circular if $M_{\rm mag}/M_{\rm BE}$ is small. The magnetic field strength of CB81 is weak, but the polarization vectors shown in Figures 6 and 7 are relatively well aligned along the distorted field, and the magnetic fields appear to be strong enough to ensure alignment. Li, McKee \& Klein (2015) conducted magnetohydrodynamic (MHD) simulations and concluded that, in the slightly magnetically supercritical case ($\lambda = 1.62$), magnetic fields are well aligned. In contrast, in the very magnetically supercritical case ($\lambda=16.2$), magnetic field lines are highly tangled by the turbulent motions, and this does not match observations. 
Comparing these two scenarios, CB81 is closer to the former case involving moderately magnetically supercritical conditions ($\lambda = 4.04$), and the relatively good alignment of magnetic fields in CB81 is consistent with the theoretical simulation. 
\par
The origin of the spatial offset of $92''$ between the center of the magnetic field configuration and the $A_V$ distribution is presently uncertain. This offset may result from the initial core formation conditions. Because the magnetic field around the core was found to be relatively uniform, it is reasonable to assume that that the initial magnetic field distribution during core formation was also uniform. The density of the medium surrounding CB81 is likely $\sim 10^3$ cm$^{-3}$ ($P_{\rm ext}/T_{\rm kin}$ for the core, see Appendix), and the magnetic fields should be frozen in the medium at this density. In this case, the initial mass distribution would not expected to be uniform, to account for the observed offset hourglass-shaped magnetic fields. In the absence of external forces such as turbulence or shocks, the center of mass for the medium and the center of the distorted magnetic fields should be in the same position, because the mass will move toward the common center of gravity and the frozen-in magnetic fields will be dragged by the medium. Thus, turbulence and/or shocks are required to migrate the medium toward a position different from the center of gravity. In this simple scenario, an external force such as turbulence or shocks plays a role in core formation, and a core formation scenario driven solely by self-gravity can be rejected. Therefore, the existence of offset hourglass-shaped fields may restrict the formation scenario of dense cores. 
%

\subsection{Polarization--Extinction Relationship}
As described in a previous paper (Paper III), the relationship between the dichroic polarization of dust ($P$) and extinction ($A$) in dark clouds is important. This relationship affects the observational interpretation of the interstellar polarization angle, which in turn is closely related to the plane-of-sky magnetic field direction, and also the investigation and restriction of dust grain alignment models. The former point is essential to guarantee that our polarization observations can trace the dust alignment and magnetic fields deep inside the CB81 core, while the latter point is important for the comparison of observations with dust alignment theories. 
\par
The observed polarizations toward CB81 represent superpositions of the polarization from the core and from the ambient medium surrounding the core. The polarization arising from the ambient off-core medium should therefore be subtracted from the observed polarization in order to isolate the polarization associated with the core. As described in Section 3.3, distorted magnetic fields surrounding the core can also produce a depolarization effect. Furthermore, the line-of-sight inclination angle in magnetic axis can affect observed plane-of-sky polarizations. These effects should be corrected in order to accurately ascertain the relationship between polarization and extinction in dense cores. 
\par
Figure 12(a) shows the observed $P_H$ vs. $H-K_{s}$ relationship with no correction (original data). The polarization is seen to generally increase with increase in the extinction value up to $H-K_{s} \sim 0.6$ mag, but exhibits a highly scattered distribution above $H-K_{s} \sim 0.6$ mag. The correlation coefficient of the relationship is only 0.49. 
The observed polarization is the superposition of the polarization from the core itself and ambient polarization. The off-core stars located outside the core radius, $R>177''$, can be used to estimate off-core polarization vectors to be subtracted from the observed polarizations toward the core (see Section 3.1), and a relatively linear $P_H$ vs. $H-K_{s}$ relationship was obtained after subtracting ambient polarization components as shown in Figure 12(b). The slope of the relationship is $1.90 \pm 0.07$ $\%$ ${\rm mag}^{-1}$. 
\par
CB81 is associated with inclined distorted magnetic fields, as discussed in Sections 3.1-3.3, which provides depolarization effects inside the core. Based on the known 3D magnetic field structure, a depolarization correction factor was estimated, as shown in Figure 13. This figure was created simply by dividing the polarization degree map of the distorted field model with edge-on geometry ($\gamma_{\rm mag} = 0^{\circ}$) by the inclined distorted field model ($\gamma_{\rm mag} = 20^{\circ}$). The correction factor map can thus simultaneously correct for both the depolarization and line-of-sight inclination angles. In Figure 13, the factors distributed around the equatorial plane are less than unity (indicating depolarization). This is due to the crossing of the polarization vectors at the front and back sides of the core along the line-of-sight (see the explanatory illustration in Figure 7 of Kataoka et al. 2012). Because the center of mass and the center of the distorted magnetic fields are different in CB81, this slightly breaks the symmetry of the distribution of depolarization correction factors. 
\par
Figure 12(c) shows the depolarization and inclination corrected $P$--$A$ relationship obtained by dividing the Figure 12(b) relationship by the correction factor map (Figure 13). In Figure 12(c), the $P$--$A$ relationship is steeper than in Figure 12(b), reflecting the correction. The slope of the relationship is $3.16 \pm 0.09$ $\%$ ${\rm mag}^{-1}$. 
\par
The correlation coefficients obtained in these analyses were 0.82 (ambient subtraction, Figure 12(b)) and 0.81 (depolarization and inclination correction, Figure 12(c)). Both are higher than the value of 0.49 determined for the original relationship. 
\par
Comparing Figures 12(a) and 12(c), the changes in both scatter and slope are dramatic. A relatively linear relationship between polarization and dust extinction in the range up to $A_V \sim 30$ mag was obtained. This result confirms that our NIR polarimetric observations accurately trace the overall polarization (magnetic field) structure of CB81. 
\par
Figure 14 presents the $P_H / A_V$ versus $A_V$ diagram. The corrected $P_H$ data as shown in Figure 12(c) was used to obtain $P_H / A_V$. Here, the dashed line indicates the linear least-squares fit to the data points with $A_V > 5$ mag, resulting in $-0.0022 A_V + 0.1959$. The linear relationship in Figure 12(c) is reflected in the shallow slope of the linear fitting result. The dotted line shows the fitting of the entire data set using the power law $P_H / A_V \propto A_V^{\alpha}$, giving an $\alpha$ index of $-0.47 \pm 0.10$. The power-law fitting appears to be identical to the linear fitting for $A_V > 5$ mag. The dotted-dashed line presents the observational upper limit reported by Jones (1989). The relationship was calculated based on the equation $P_{K,{\rm max}} = \tanh{\tau_{\rm p}}$, where $\tau_{\rm p} = (1-\eta)\tau_{K}/(1+\eta)$ and the parameter $\eta$ is set to 0.875 (Jones 1989). $\tau_{K}$ denotes the optical depth in the $K$ band and $P_H / A_V \approx 0.62$ at $\tau_{K} = 1$. 
\par
The polarization efficiency of CB81 is about half that of FeSt 1-457 (Paper VI). CB81 and FeSt 1-457 are associated with the same dark cloud complex (Pipe Nebula), and this difference in efficiency implies that the dust properties and/or the radiation environment (based on radiative torque grain alignment theory: Dolginov \& Mitrofanov 1976; Draine \& Weingartner 1996,1997; Lazarian \& Hoang 2007) can vary even in the same dark cloud complex. 

\section{Summary and Conclusion}
The present study ascertained the detailed magnetic field structure of the starless dense core CB81 (L1774, Pipe 42) based on NIR polarimetric observations of background stars to measure dichroically polarized light produced by aligned dust grains in the core. After subtracting ambient polarization components, the magnetic fields pervading CB81 were mapped using 147 stars, and axisymmetrically distorted hourglass-like magnetic fields were identified. On the basis of simple 2D and 3D magnetic field modeling, magnetic inclination angles relative to the plane-of-sky and line-of-sight directions were determined to be $4^{\circ} \pm 8^{\circ}$ and $20^{\circ} \pm 20^{\circ}$, respectively. Using these angles and the Davis-Chandrasekhar-Fermi method, the total magnetic field strength of CB81 was found to be $B_{\rm pos}/ \sin(\theta_{\rm inc}) = 6.8/ \cos(20) = 7.2 \pm 2.3$ $\mu$G. The magnetic critical mass of the core, $M_{\rm mag}=0.83 \pm 0.25$ M$_{\odot}$, was determined to be less than the observed core mass, $M_{\rm core}=3.37 \pm 0.51$ M$_{\odot}$, suggesting a magnetically supercritical state with the ratio of observed mass-to-flux ratio to a critical value, $\lambda = 4.04 \pm 0.80$. The critical mass of CB81, evaluated by incorporating both magnetic and thermal+turbulent support effects, $M_{\rm cr} \simeq M_{\rm mag}+M_{\rm BE}$, was determined to be $0.83+3.20=4.03 \pm 0.40$ ${\rm M}_{\odot}$, where $3.20 \pm 0.32$ ${\rm M}_{\odot}$ is the Bonnor--Ebert mass. Although the resulting $M_{\rm cr}$ value is slightly greater than the core mass, $M_{\rm core}$ (marginally stable), we conclude that CB81 is close to the critical state, where $M_{\rm cr} \sim M_{\rm core}$. A spatial offset of $92''$ was determined between the center of the magnetic field configuration and the $A_V$ distribution, possibly due to the initial conditions related to the formation of the core. Because the magnetic field around the core is fairly uniform, a simple interpretation is that the initial mass distribution during core formation was not uniform but rather biased, and this configuration was subsequently compressed by turbulence or shocks to create the observed offset hourglass-shaped magnetic fields. Self-gravity cannot drive such a process, and so the existence of these offset hourglass-shaped magnetic fields suggests a structure that restricts the core formation process. We obtained a linear relationship between the polarization and extinction values, up to $A_V \sim 30$ mag toward the stars with deepest obscuration. The slope value obtained from this relationship, $3.16 \pm 0.09$ \% mag$^{-1}$, is relatively small. This result can possibly be ascribed to a lack of a large grain population or the absence of a strong radiation field, or a combination of both effects. The linear relationship indicates that the observed polarizations reflect the overall magnetic field structure of the core. Further theoretical and observational studies would be desirable with regard to explaining the dust alignment in the dense core environment. 

\bigskip

We are grateful to the staff of SAAO for their kind help during the observations. We wish to thank Tetsuo Nishino, Chie Nagashima, and Noboru Ebizuka for their support in the development of SIRPOL, its calibration, and its stable operation with the IRSF telescope. The IRSF/SIRPOL project was initiated and supported by Nagoya University, National Astronomical Observatory of Japan, and the University of Tokyo in collaboration with South African Astronomical Observatory under the financial support of Grants-in-Aid for Scientific Research on Priority Area (A) No. 10147207 and No. 10147214, and Grants-in-Aid No. 13573001 and No. 16340061 of the Ministry of Education, Culture, Sports, Science, and Technology of Japan. RK, MT, NK, KT (Kohji Tomisaka), and MS also acknowledge support by additional Grants-in-Aid Nos. 16077101, 16077204, 16340061, 21740147, 26800111, 16K13791, 15K05032, 16K05303, 19K03922.

\bigskip

\subsection*{A1: Dust Extinction Map and the Bonnor--Ebert Model Fitting}
Dust extinction (i.e., $A_V$) measurement, particularly at NIR wavelengths, is one of the most straightforward methods for revealing the density structure within dense dark clouds and cores. There are two approaches to these measurements: the star count method (e.g., Wolf 1923; Dickman 1978; Cambr\'{e}sy 1999; Dobashi et al. 2005) based on determining the stellar density distribution and the NIR color excess (NICE) method (Lada et al. 1994; NICER by Lombardi \& Alves 2001; NICEST by Lombardi 2009; XNICER by Lombardi 2018) based on the measurement of the stellar color excess. Considering the stellar densities at the $H$ and $K_s$ bands in the densest region of CB81, we elected to use the latter approach, because the dust extinction toward the core center was not severe and we could detect many reddened background stars through the core. 
\par
The extinction map was produced following the procedure described in Kandori et al. (2005). Assuming that the population of stars across the CB81 field is invariable, we were able to use the mean stellar color in the reference field, $<H-K_s>_{\rm ref}$, as the zero point for color excess in the CB81 core. This reference field is situated at the north-west and south-east edges of the observation field. The color excess distribution of the core was obtained by subtracting $<H-K_s>_{\rm ref}$ from the observed $H-K_s$ color of stars. Since the extent of the analyzed field was relatively small ($\sim 8' \times 8'$), our assumption of an invariant stellar population is considered to be reasonable. The color excess distribution was determined by arranging circular cells (30$''$ in diameter) spaced 9$''$ apart on the observation field. The mean stellar color in each cell was calculated and the color excess as derived according to the equation 
\begin{equation}
{E}_{H-K_s}\left({ \Delta \alpha , \Delta \delta }\right)=\left[{ \sum\limits_{ i=1}^{ N} {\frac{ (H-K_s{)}_{i}}{N}}}\right]\left({ \Delta \alpha , \Delta \delta }\right)-{\left\langle{H-K_s}\right\rangle}_{\rm ref}, 
\end{equation}
where $(H-K_s){}_{i}$ is the color index of the $i$-th star in a cell, $N$ is the number of stars in a cell, and $\langle H-K_s \rangle {}_{\rm ref}$ is the mean color of stars in the reference field. The $H-K_s$ color excess was converted to $A_{V}$ using $A_{V}=21.7\times E_{H-K_s}$ (Nishiyama et al. 2008), following which the map was smoothed with a FWHM=15$''$ Gaussian filter. The resulting $A_{V}$ map is shown in Figure S1. The typical uncertainty associated with $A_{V}$ in the reference field ($A_{V} \sim 0$ mag) was estimated to be $\sim0.5$ mag. The position of the core center based on the centroid of the $A_V$ distribution is R.A.=17$^{\rm h}$22$^{\rm m}$41$.\hspace{-3pt}^{\rm s}$42, Decl.=-27$^{\circ}$05$'$12$.\hspace{-3pt}''8$ (J2000). 
\par
Following the procedures detailed in Sections 4.1 and 4.2 of Kandori et al. (2005), we produced a radial column density profile for CB81 and conducted the fitting using the Bonnor--Ebert model (Ebert 1955; Bonnor 1956). Note that we manually masked extinction features likely to be unrelated to CB81 (i.e., a diffuse streaming feature to the north-west in Figure S1) and masked regions were ignored in the circular averaging procedure. Figure S2 presents the results of the Bonnor--Ebert fitting of CB81. Here, the dots and error bars represent the average $N$(H${}_{2}$) values at each annulus at 4$.\hspace{-3pt}''$5 intervals and the root mean square (rms) deviation of data points in each annulus, respectively. The solid line denotes the radial column density profile of the best-fit Bonnor--Ebert sphere convolved with beam used in the $A_{V}$ measurements. The dot-dashed line indicates the Bonnor--Ebert model profile before the convolution. 
\subsection*{A2: Physical Properties of CB81}
The physical properties we obtained for the CB81 core include a radius, $R$, of $23,000 \pm 1,000$ AU ($177'' \pm 7''$), central density, $\rho_{\rm c}$, of $9.0(\pm 2.1) \times 10^{-19}$ g cm$^{-3}$ $(2.31 (\pm 0.54) \times 10^{5}$ cm$^{-3}$), temperature, $T$, of $15.0 \pm 0.8$ K, mass, $M_{\rm core}$, of $3.37 \pm 0.51$ M$_{\odot}$, external pressure, $P_{\rm ext}$, of $4.33(\pm 0.75) \times 10^4$ K cm$^{-3}$ and a Bonnor--Ebert stability parameter, $\xi_{\rm max}$, of $13.0 \pm 1.7$. Note that the temperature value was determined from the Bonnor--Ebert fitting, and is roughly consistent with the \lq \lq effective'' temperature of CB81 of $T_{\rm kin} + T_{\rm turb} = T_{\rm kin} + \sigma_{\rm turb}^2 m/k = 12.4 + 1.11 = 13.51$ K, where $T_{\rm kin}$ is the kinematic temperature measured using NH$_3$ lines and $T_{\rm turb}$ is the temperature calculated using a turbulent velocity dispersion of 0.0627 km s$^{-1}$ (Rathborne et al. 2008). Because the distance, $d$, and temperature, $T$, are coupled in the Bonnor--Ebert solution as $d^{-1}T={\rm constant}$ (Lai et al. 2003), the consistency of $T$ values obtained from the Bonnor--Ebert fitting and from the independent radio measurements indicates that the distance measured by Lombardi et al. (2006) is reasonably accurate. 
\par
The starless dense core CB81 is now known to be a Bonnor--Ebert core with $\xi_{\rm max} \sim 13.0$. Because its critical value exceeds $\xi_{\rm max}=6.5$, CB81 is considered to be unstable with respect to Bonnor--Ebert equilibrium. As shown by Kandori et al. (2005), a series of solutions indicating an unstable Bonnor--Ebert equilibrium is indistinguishable from the density structure evolution of a collapsing sphere. Therefore, CB81 could be undergoing gravitational collapse if the core does not contain additional supporting force acting against gravity. A survey looking for gas infalling motion using the HCN ($J=1-0$) molecular line failed to detect gas inward motion toward CB81 because the emission from the core is too faint (Afonso et al. 1998). To confirm the line-of-sight gas motion toward CB81, more sensitive radio observations are needed. It is obvious that CB81 core has additional support from magnetic fields. With the support, the core can be in kinematically nearly critical condition. Kandori et al. (2005) suggested that the majority of starless cores show a nearly critical Bonnor--Ebert solution. Because CB81 is starless and has a steep density structure, as indicated by $\xi_{\rm max} \sim 13.0$, it can serve as an important object for the study of star formation by illustrating conditions just before or after the onset of gravitational collapse. 

\subsection*{A3: Accuracy of Flat Fielding}
Here, we discuss the accuracy of flat fielding associated with calibration of the SIRPOL data. At present, SIRPOL data is calibrated using a flat image based on data acquired toward the twilight sky (i.e., twilight flat, see Figure S3). Twilight sky images with different sky count levels can be used to generate pairs of images, following which differential images generated from each pair are combined and normalized to obtain the final twilight flat image. During the exposures to acquire these twilight flat images, the angle of the half-waveplate is fixed at $67.5^{\circ}$. 
\par
There are two issues that may affect the accuracy of the SIRPOL flat image: (1) the sky is generally polarized, and this may affect the accuracy of SIRPOL flat, and (2) it is also uncertain whether the twilight flat image created based on the fixed waveplate angle ($67.5^{\circ}$) is applicable to the calibration of other waveplate angle images. 
\par
To ascertain these issues, we employed sky images obtained during the observations of CB81. A total of 10 sets of sky images (that is, 10 exposures repeated 10 times) were used for each waveplate angle image, and Figure S4 shows the Stokes I image of the median sky in the $H$ band. Here, all the sky images generated at the four waveplate angles are combined. The vertical gap in the sky count at $x=512$ pixels is termed the \lq \lq reset anomaly pattern.'' The median sky count of the image is $1408$ ADU. Since the sky count in the $H$ band is due to the OH airglow lines, we believe that the sky was not polarized. The effect of the moon (illumination=66\%, distance=$74^{\circ}$ on June 15 2017 and dark on June 24 2017) can be used to set the upper limit for the analysis, as described below. 
\par
We evaluated the fraction of polarized light using the median sky images of $I_{00}$, $I_{22.5}$, $I_{45}$, and $I_{67.5}$, where $I_{\rm angle}$ is the image acquired with the waveplate angles of $0^{\circ}$, $22.5^{\circ}$, $45^{\circ}$, and $67.5^{\circ}$. If the SIRPOL flat image can generate an artificial pattern on calibrated images, the effect obtained from the flat image (i.e., an artificial polarization pattern) will be observed in the $H$ band median sky image. To remove small scale variations, the median sky images were smoothed using a median filter with a width of 9 pixels. 
\par
The white vectors in Figure S4 show the polarization vector distribution of the median sky. During calculations of these polarization vectors, a uniform component was subtracted. Specifically, $Q_{\rm corr}=Q-{\rm median}(Q)$ and $U_{\rm corr}=U-{\rm median}(U)$ were used because the subtraction analysis of uniform vector components is employed in our basic procedure for detecting hourglass-shaped magnetic fields. 
The values of ${\rm median}(Q)$ and ${\rm median}(U)$ are $2.07$ ADU and $2.26$ ADU, respectively. 
This subtraction analysis suppresses the resulting polarization degree by $\sim 50\%$. Figure S5 presents a histogram of the polarization degrees, for which the median and standard deviation values are $0.13\% \pm 0.08\%$. This value is less than the polarization degree threshold for the analysis of an hourglass-shaped field ($P_H \ge 0.5\%$). Moreover, the polarization angle distribution does not resemble an hourglass. Therefore, we conclude that the use of the present SIRPOL flat image does not affect the validity of our conclusion. 

\clearpage 
\begin{figure}[t]
\begin{center}
 \includegraphics[width=6.5 in]{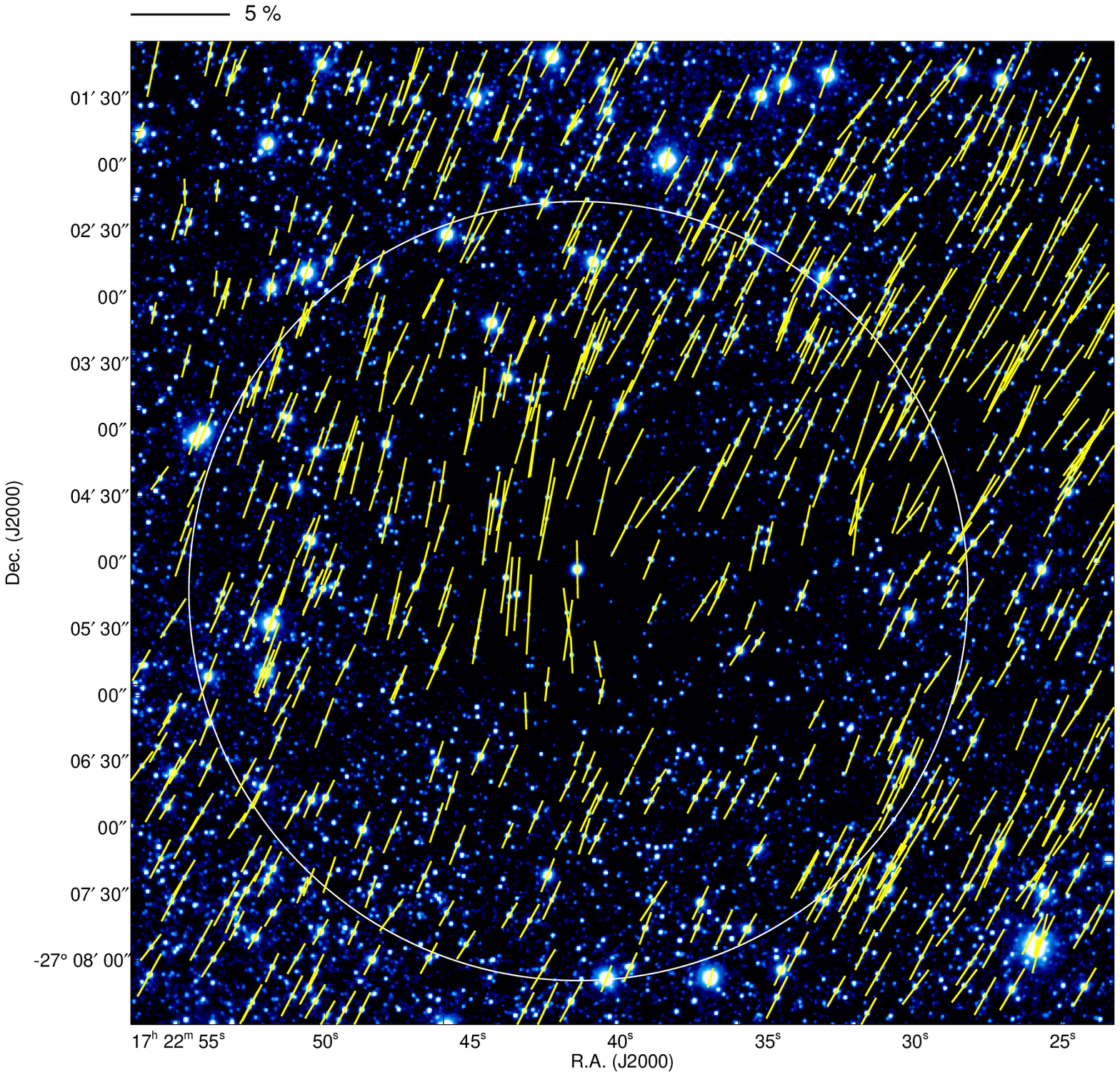}
\end{center}
 \caption{Polarization vectors of point sources superimposed on the intensity image in the $H$ band. Stars for which $P_H / \delta P_H \ge 5$ are shown. The core radius (177$''$) is indicated by the white circle. The scale bar above the image indicates 5\% polarization.}
   \label{fig1}
\end{figure}

\clearpage 
\begin{figure}[t]
\begin{center}
 \includegraphics[width=6.5 in]{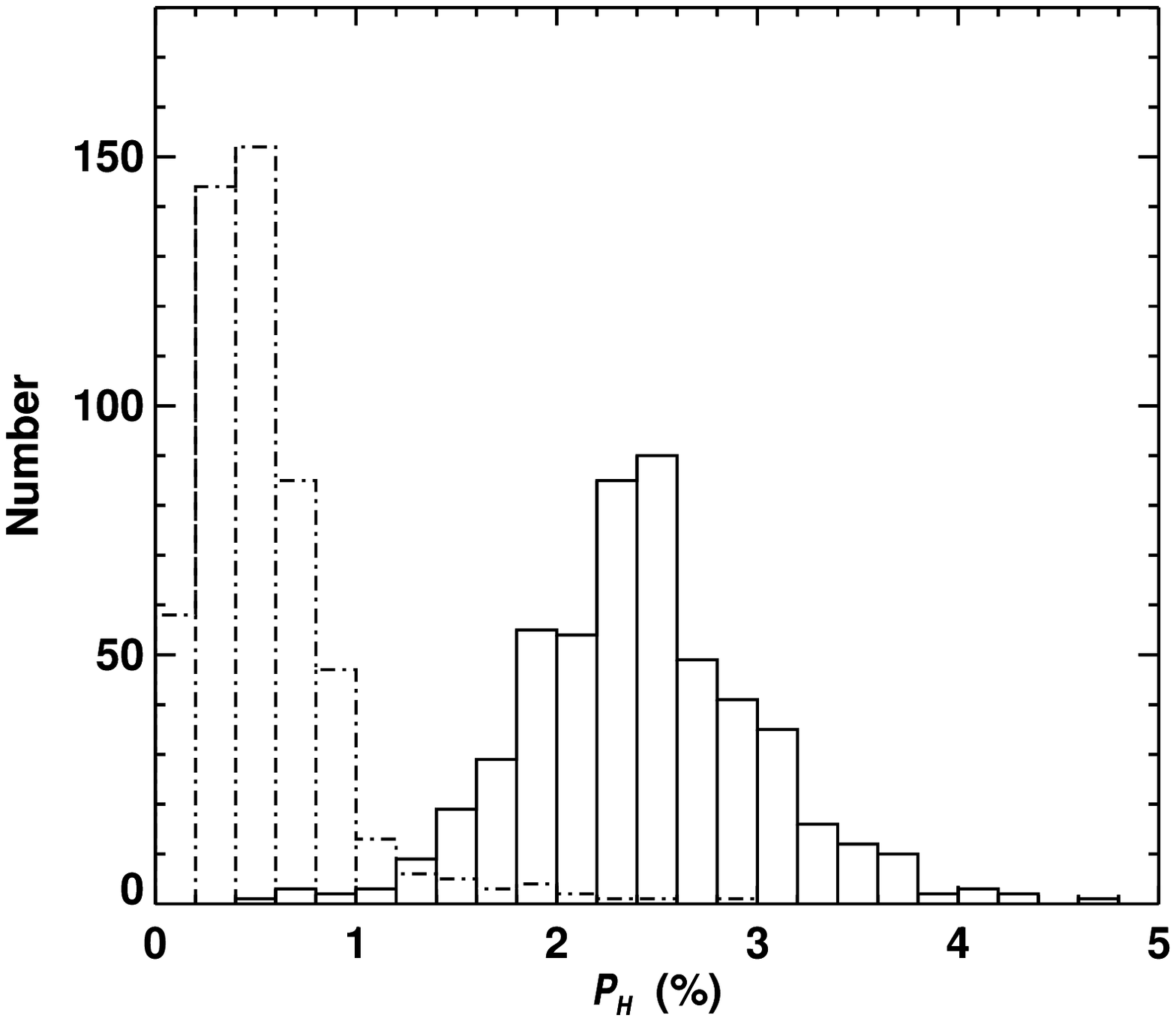}
\end{center}
 \caption{Histograms of $P_H$ values for the stars in the off-core region before (solid line) and after (dotted-dashed line) subtraction of the off-core component.}
   \label{fig1}
\end{figure}

\clearpage 
\begin{figure}[t]
\begin{center}
 \includegraphics[width=6.5 in]{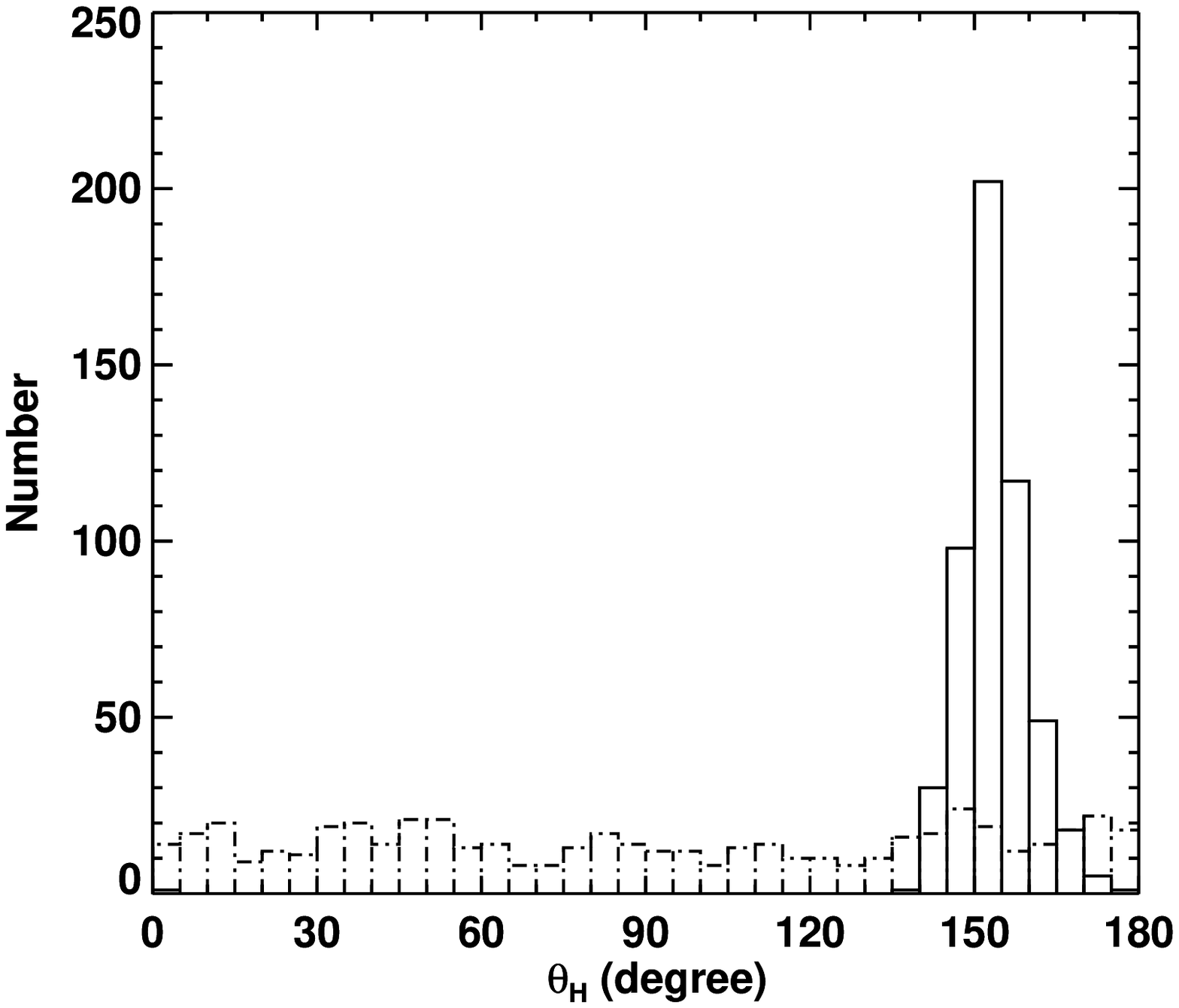}
\end{center}
 \caption{Histogram of $\theta_H$ values for the stars in the off-core region before (solid line) and after (dotted-dashed line) subtraction of the off-core component.}
   \label{fig1}
\end{figure}

\clearpage 
\begin{figure}[t]
\begin{center}
 \includegraphics[width=6.5 in]{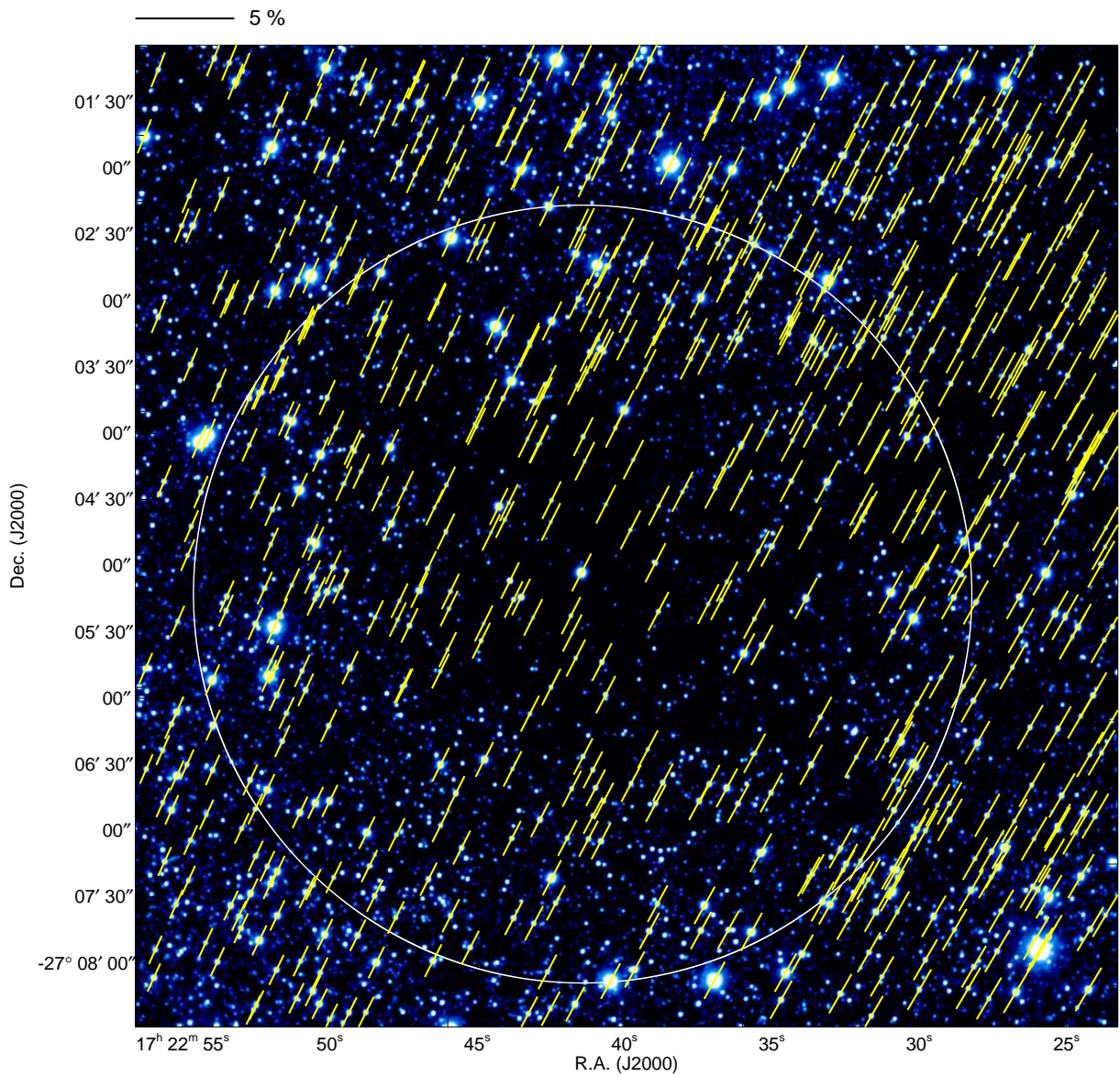}
\end{center}
 \caption{Estimated off-core polarization vectors superimposed on an intensity image in the $H$ band. Note that these vectors were not obtained directly from observations. Off-core vectors estimated by fitting are plotted at the position of each star. The scale bar above the image indicates 5\% polarization.}
   \label{fig1}
\end{figure}

\clearpage 
\begin{figure}[t]
\begin{center}
 \includegraphics[width=6.5 in]{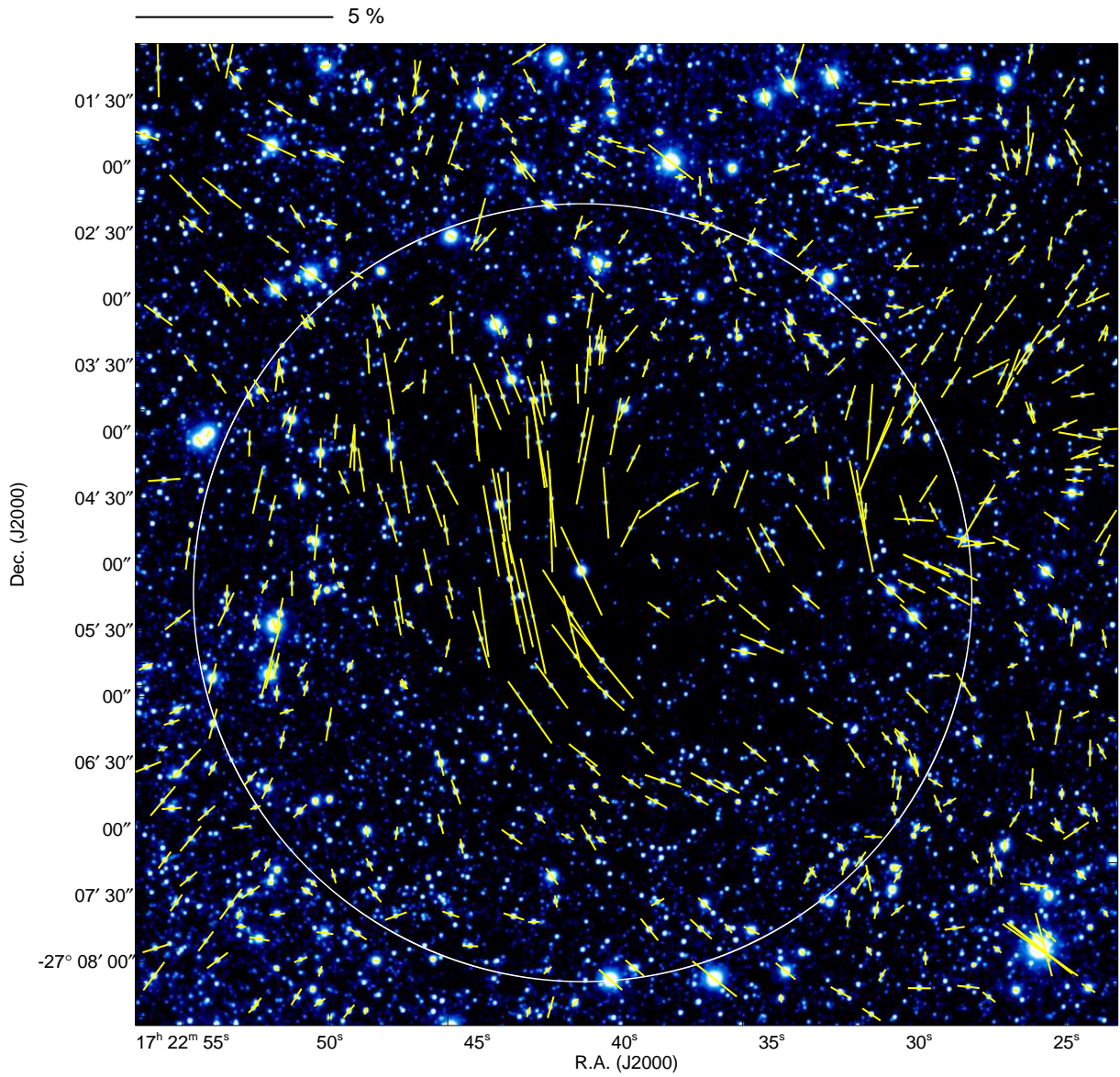}
\end{center}
 \caption{Polarization vectors after subtraction of the off-core component. The scale bar above the image indicates 5\% polarization.}
   \label{fig1}
\end{figure}

\clearpage 
\begin{figure}[t]
\begin{center}
 \includegraphics[width=6.5 in]{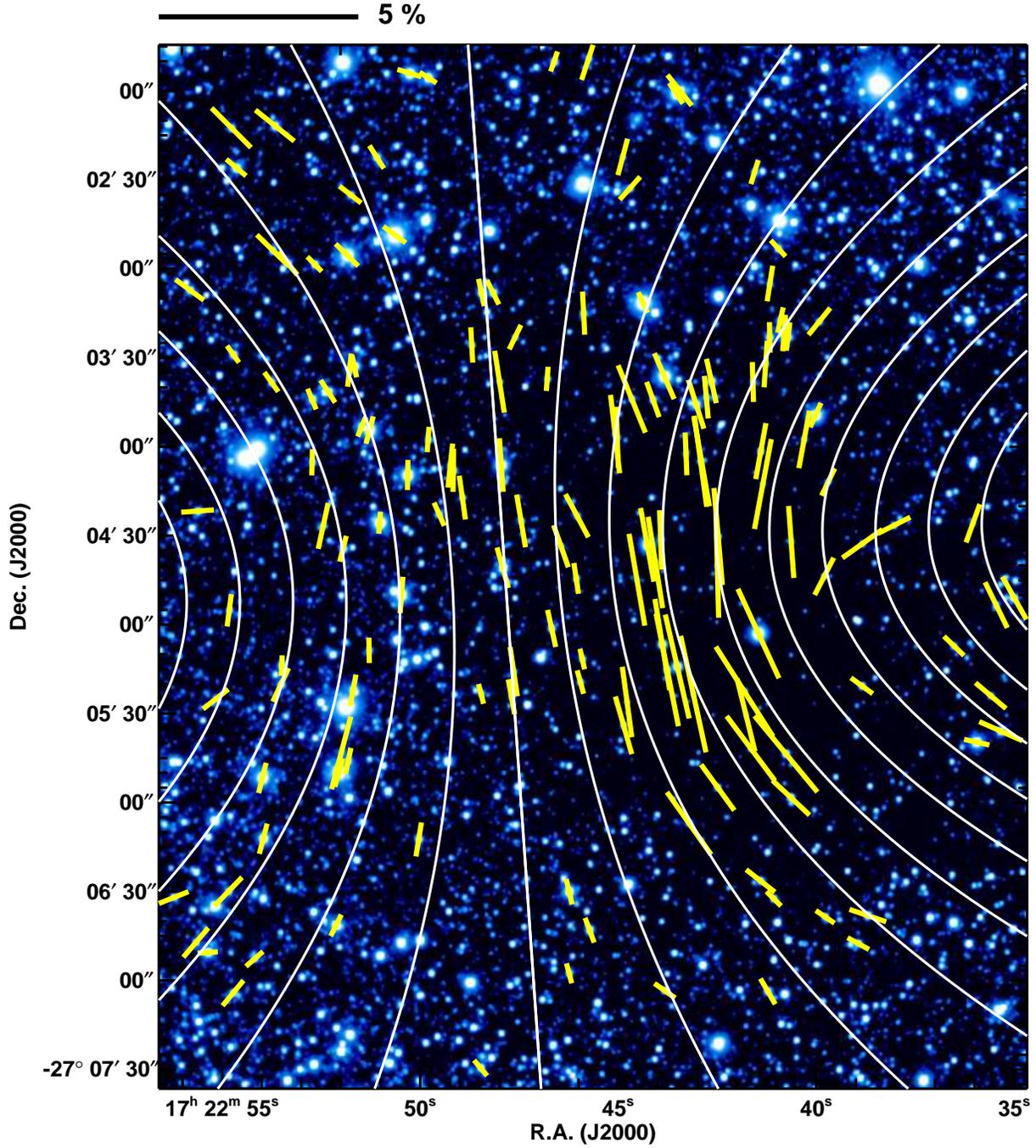}
\end{center}
 \caption{Polarization vectors after subtraction of the off-core component. The field of view is $294'' \times 354''$ ($0.19 \times 0.23$ pc) in the $\alpha$ and $\delta$ directions at a distance of 130 pc. The field size in the $\delta$ direction is equal to the diameter of the core. The background image for this figure is a magnified version of that in Figure 5. The white lines indicate the direction of the magnetic field inferred from parabolic fitting. The scale bar above the image indicates 5\% polarization.}
   \label{fig1}
\end{figure}

\clearpage 
\begin{figure}[t]
\begin{center}
 \includegraphics[width=6.5 in]{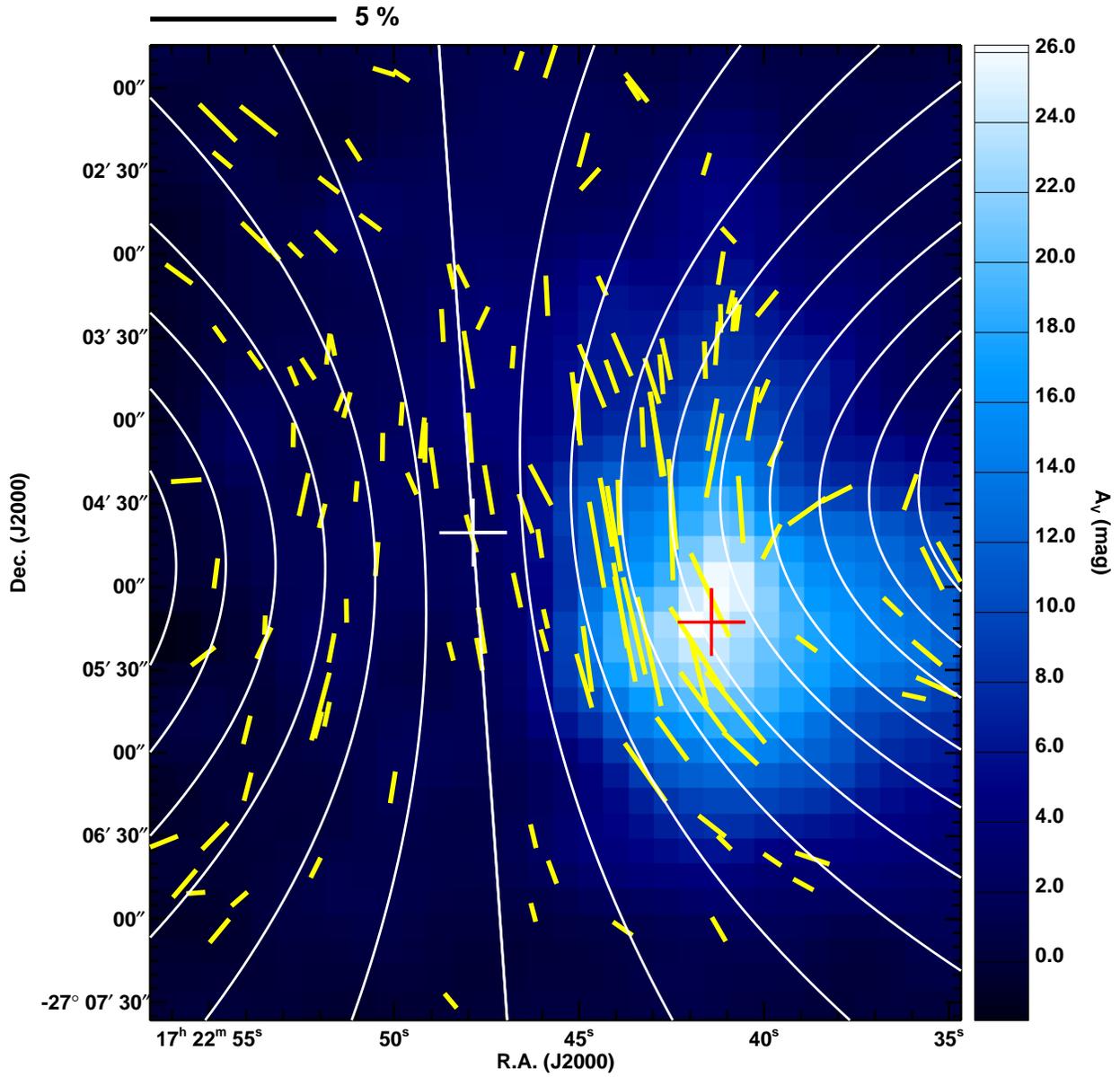}
\end{center}
 \caption{The same data as provided in Figure 6 but using the $A_V$ distribution (see Appendix) as the background image. The white and red cross indicate the center of magnetic field geometry and the centroid center of the $A_V$ distribution.}
   \label{fig1}
\end{figure}

\clearpage 
\begin{figure}[t]
\begin{center}
 \includegraphics[width=6.5 in]{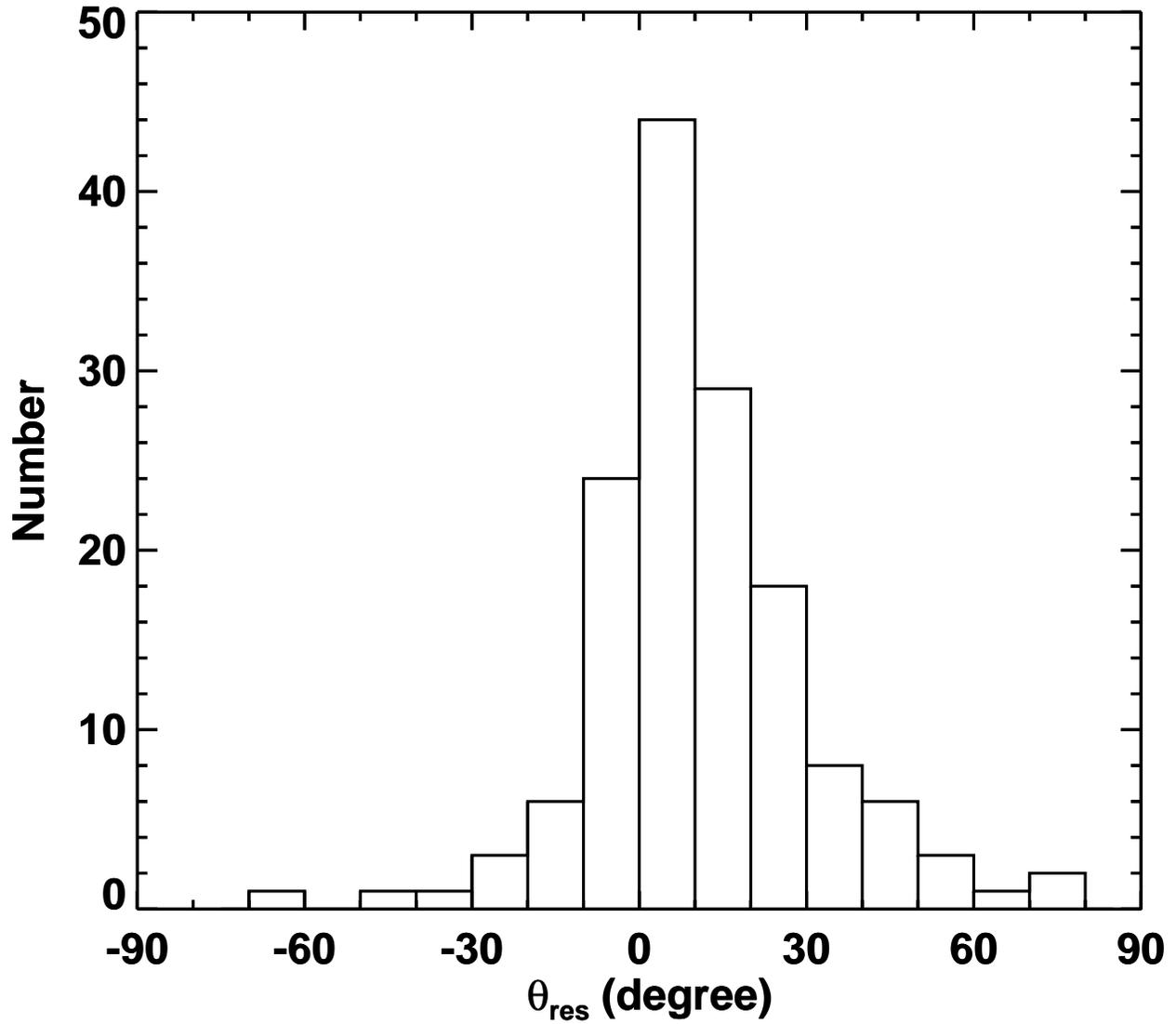}
\end{center}
 \caption{A histogram of the residuals for the observed polarization angles after subtraction of the angles obtained by parabolic fitting ($\theta_{\rm res}$).}
   \label{fig1}
\end{figure}


\clearpage 
\begin{figure}[t]
\begin{center}
 \includegraphics[width=6.5 in]{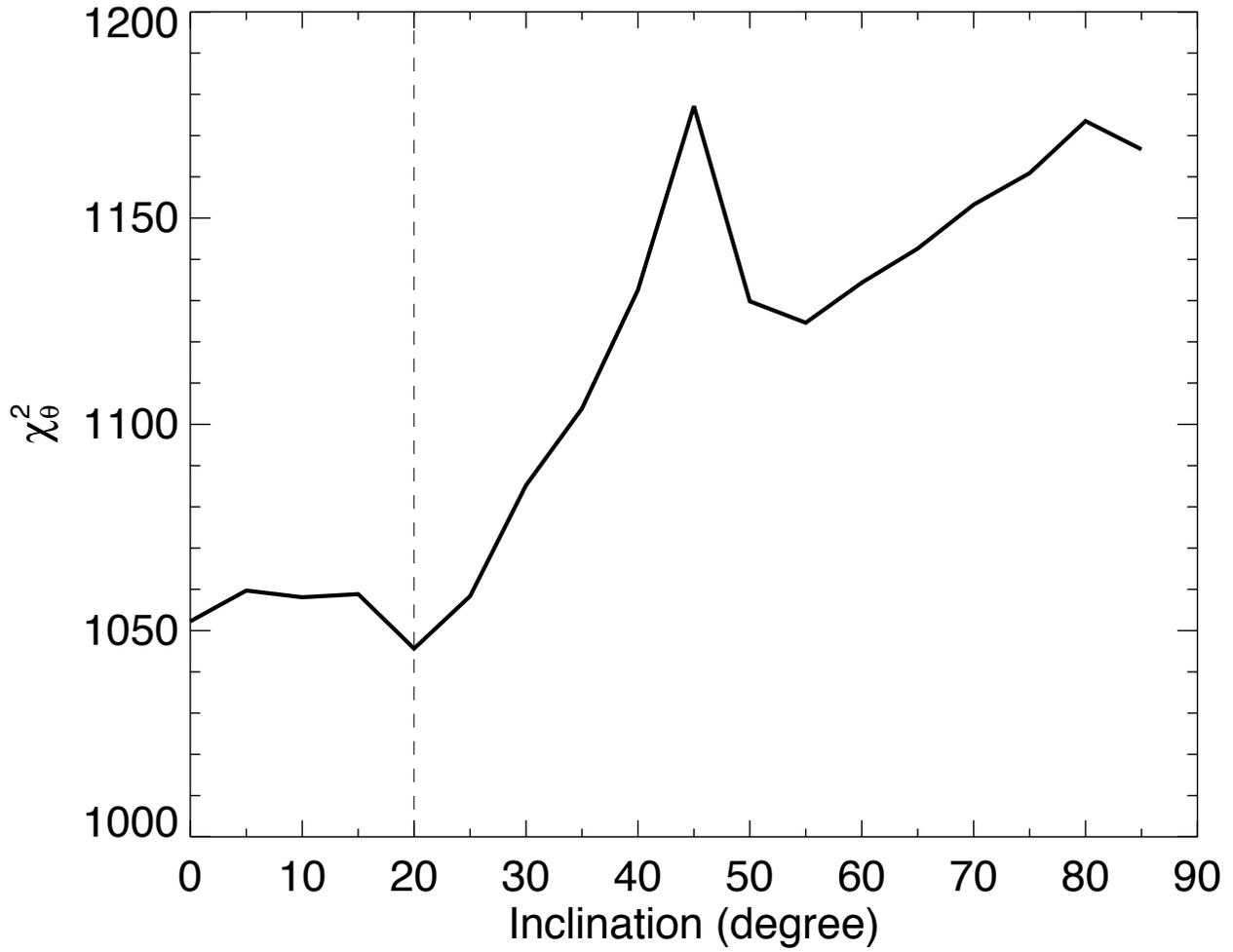}
\end{center}
 \caption{The $\chi^{2}$ distribution of the polarization angles ($\chi^{2}_{\theta}$). The best magnetic curvature parameter ($C$) was determined at each $\theta_{\rm inc}$. $\theta_{\rm inc}=0^{\circ}$ and $90^{\circ}$ correspond to the edge-on and pole-on geometry in the magnetic axis.}
   \label{fig1}
\end{figure}


\clearpage 
\begin{figure}[t]
\begin{center}
 \includegraphics[width=6.5 in]{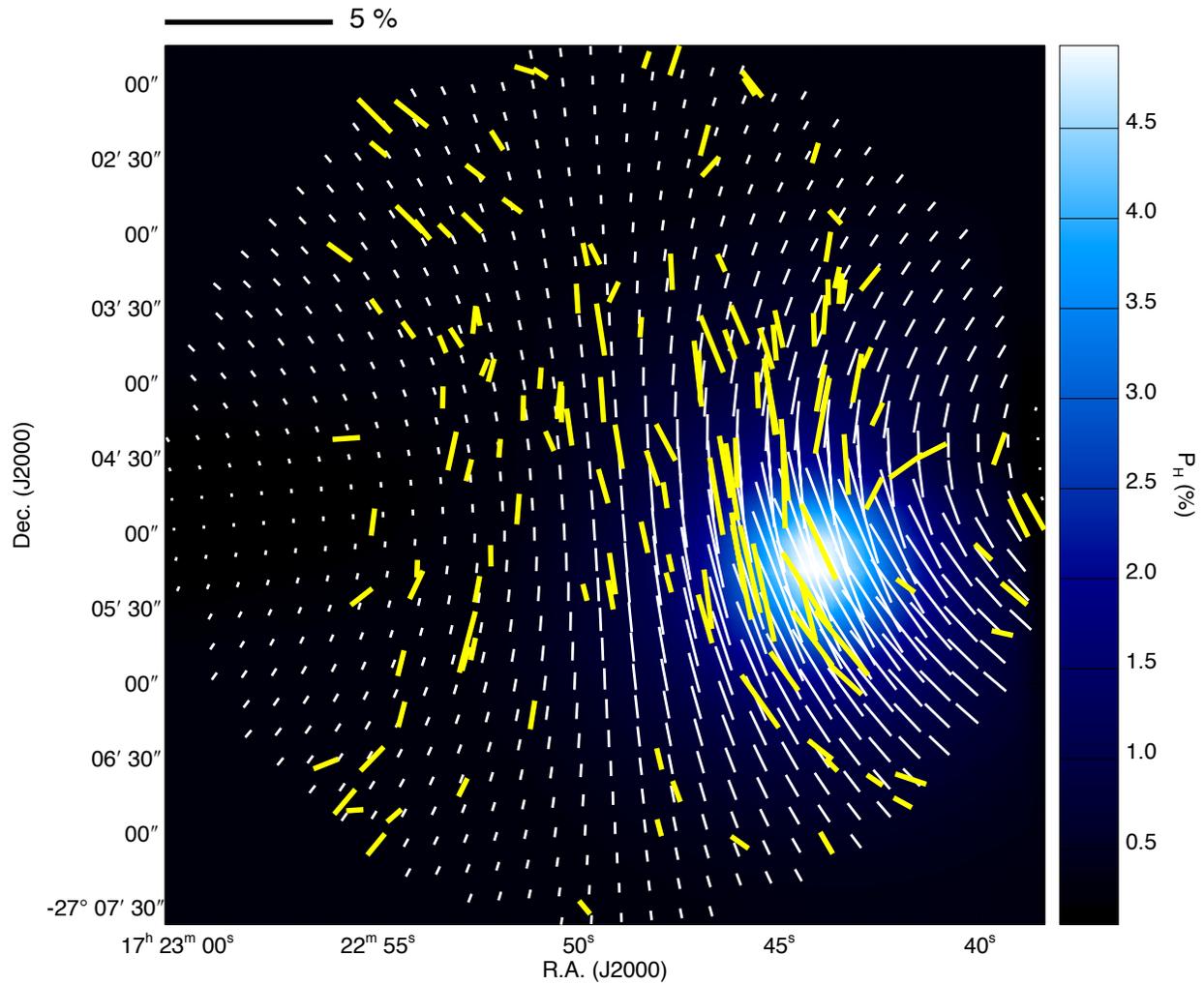}
\end{center}
 \caption{The best-fit 3D parabolic model (white vectors) with observed polarization vectors (yellow vectors). The background color image shows the polarization degree distribution of the best-fit model. The scale bar above the image indicates 5\% polarization.}
   \label{fig1}
\end{figure}

\clearpage 
\begin{figure}[t]
\begin{center}
 \includegraphics[width=6.5 in]{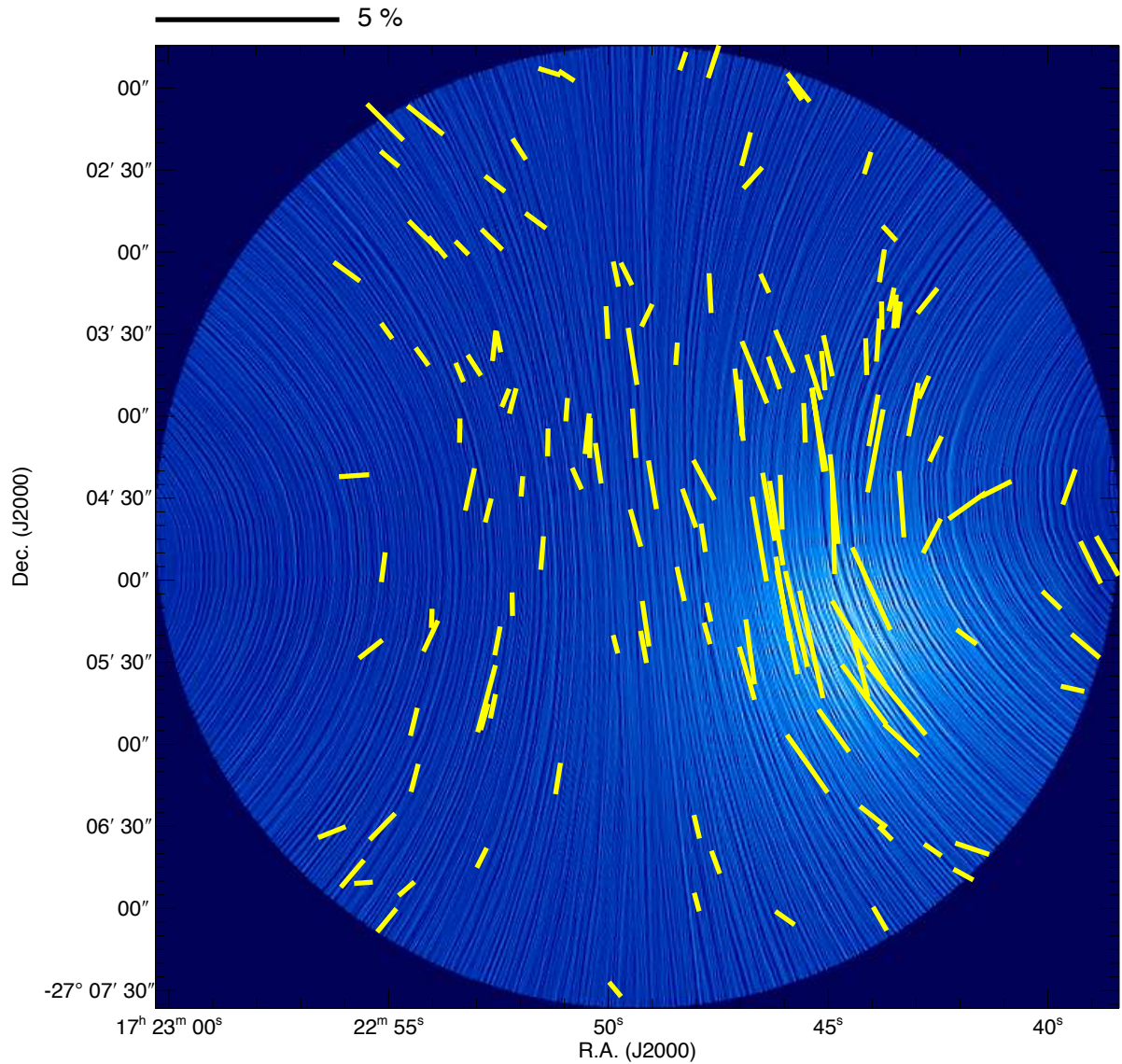}
\end{center}
 \caption{The same data as presented in Figure 10, but with the background image generated using the line integral convolution (LIC) technique (Cabral \& Leedom 1993). The direction of the LIC \lq \lq texture'' is parallel to the direction of the magnetic fields and the background image is based on the polarization degree of the model core.}
   \label{fig1}
\end{figure}

\clearpage 
\begin{figure}[t]
\begin{center}
 \includegraphics[width=2.5 in]{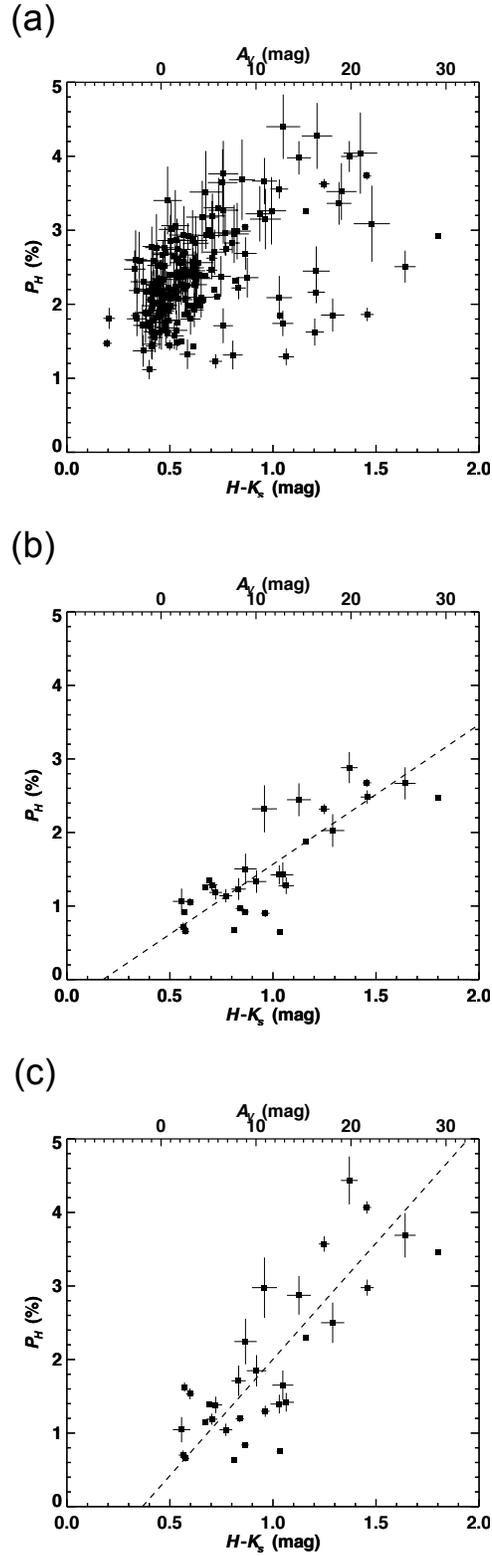}
\end{center}
 \caption{The relationship between the polarization degree and the $H-K_{\rm s}$ color toward background stars. Stars for which $R \leq 177''$ and $P/\delta P \geq 6$ are plotted. The dashed line denotes the linear fit to the data. The $P$--$A$ relationship (a) without any correction (original data), (b) after correcting for ambient polarization components, and (c) after correcting for ambient polarization components, the depolarization effect, and the magnetic inclination angle.}
   \label{fig1}
\end{figure}

\clearpage 
\begin{figure}[t]
\begin{center}
 \includegraphics[width=6.5 in]{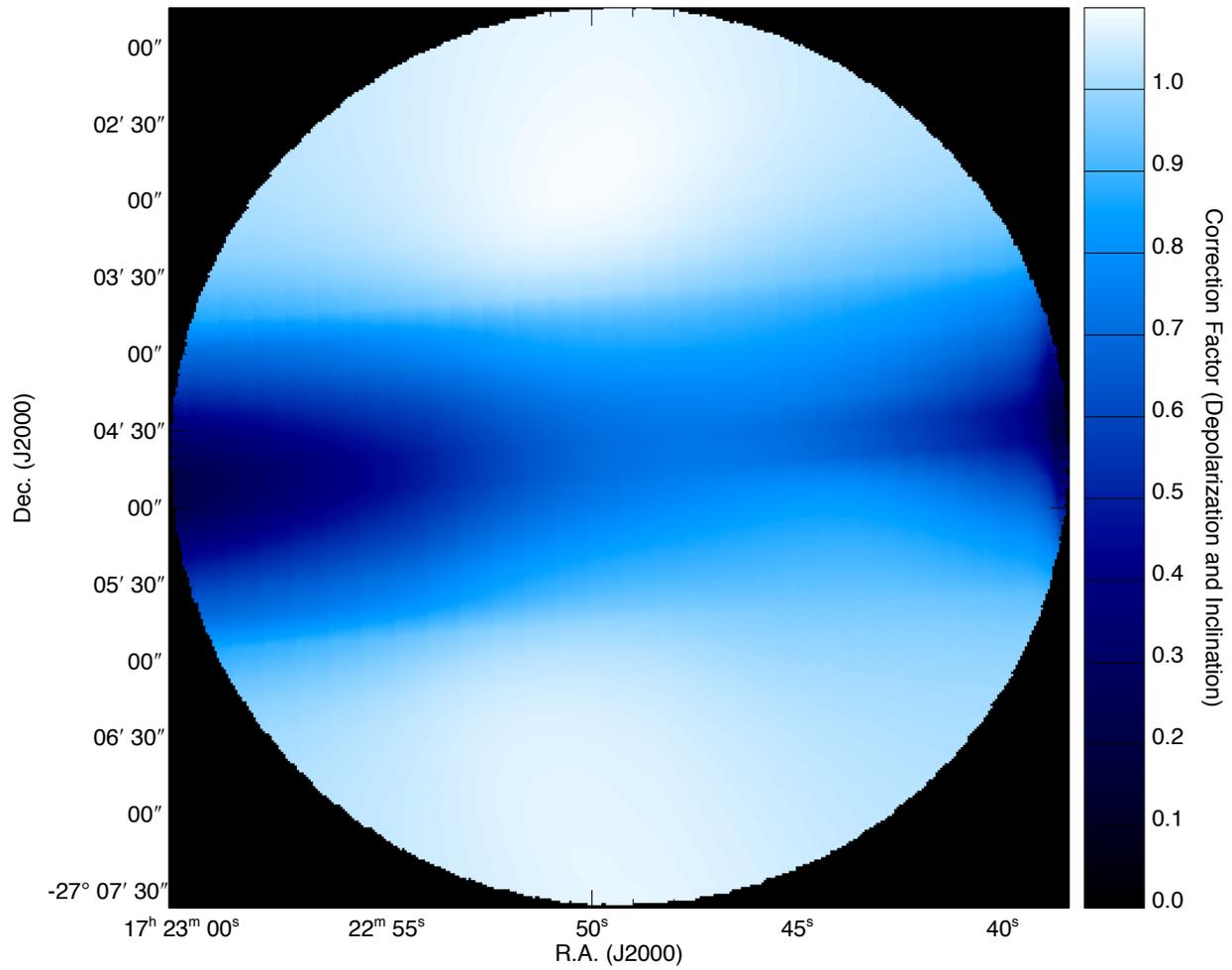}
\end{center}
 \caption{Distribution of depolarization correction factors. The field of view is the same as the diameter of the core ($354''$).}
   \label{fig1}
\end{figure}

\clearpage 
\begin{figure}[t]
\begin{center}
 \includegraphics[width=6.5 in]{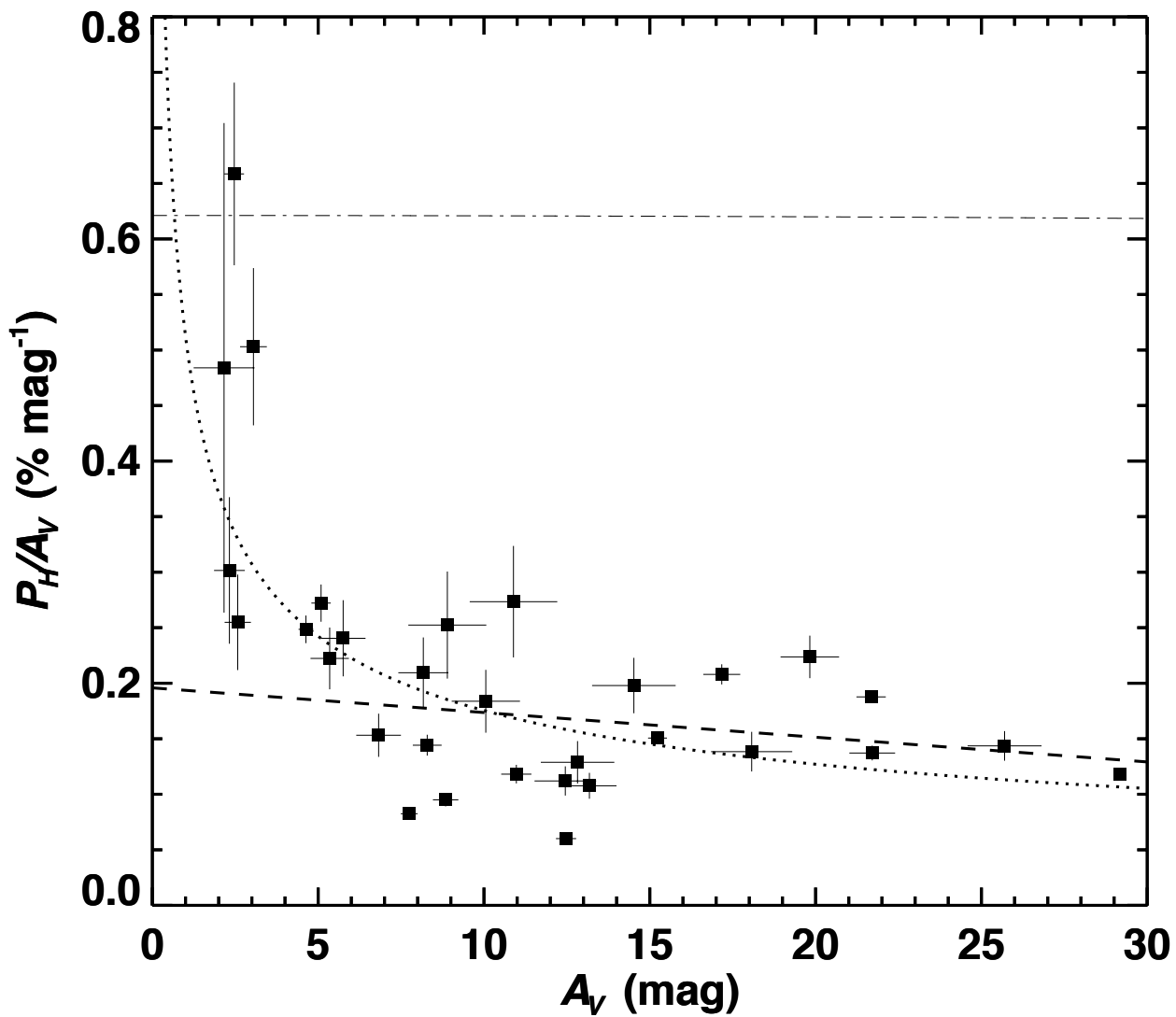}
\end{center}
 \caption{The relationship between polarization efficiency $P_H / A_V$ and $A_V$ toward the background stars of CB81. The stars for which $R \leq 177''$ and $P/\delta P \geq 6$ are plotted. The dashed line denotes the linear fit to the data with $A_V > 5$ mag. The dotted line shows the power-law fit for all the data points. The dotted-dashed line shows the observational upper limit reported by Jones (1989).}
   \label{fig1}
\end{figure}

\clearpage

\beginsupplement

\clearpage 

\begin{figure}[t]  
\begin{center}
 \includegraphics[width=6.5 in]{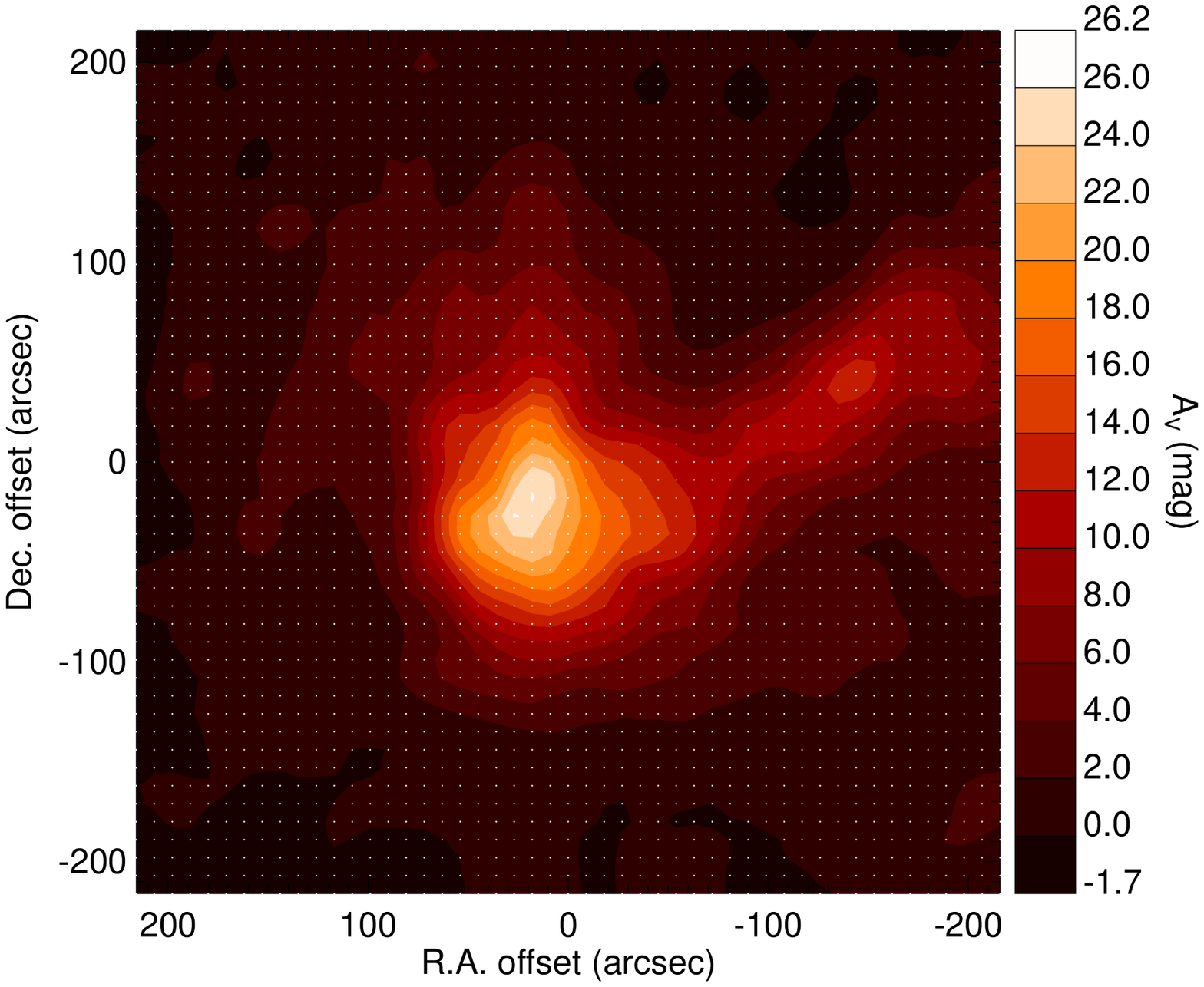}
\end{center}
 \caption{The $A_V$ distribution in the starless dense core CB81. White dots represent the $A_V$ measurement points separated by $9''$. The resolution of the map is $34''$ and the position of the field center is R.A.=17$^{\rm h}$22$^{\rm m}$39$.\hspace{-3pt}^{\rm s}$93, Decl.=-27$^{\circ}$04$'$46$.\hspace{-3pt}''5$ (J2000).}
   \label{fig5}
\end{figure}

\clearpage

\begin{figure}[t]  
\begin{center}
 \includegraphics[width=6.5 in]{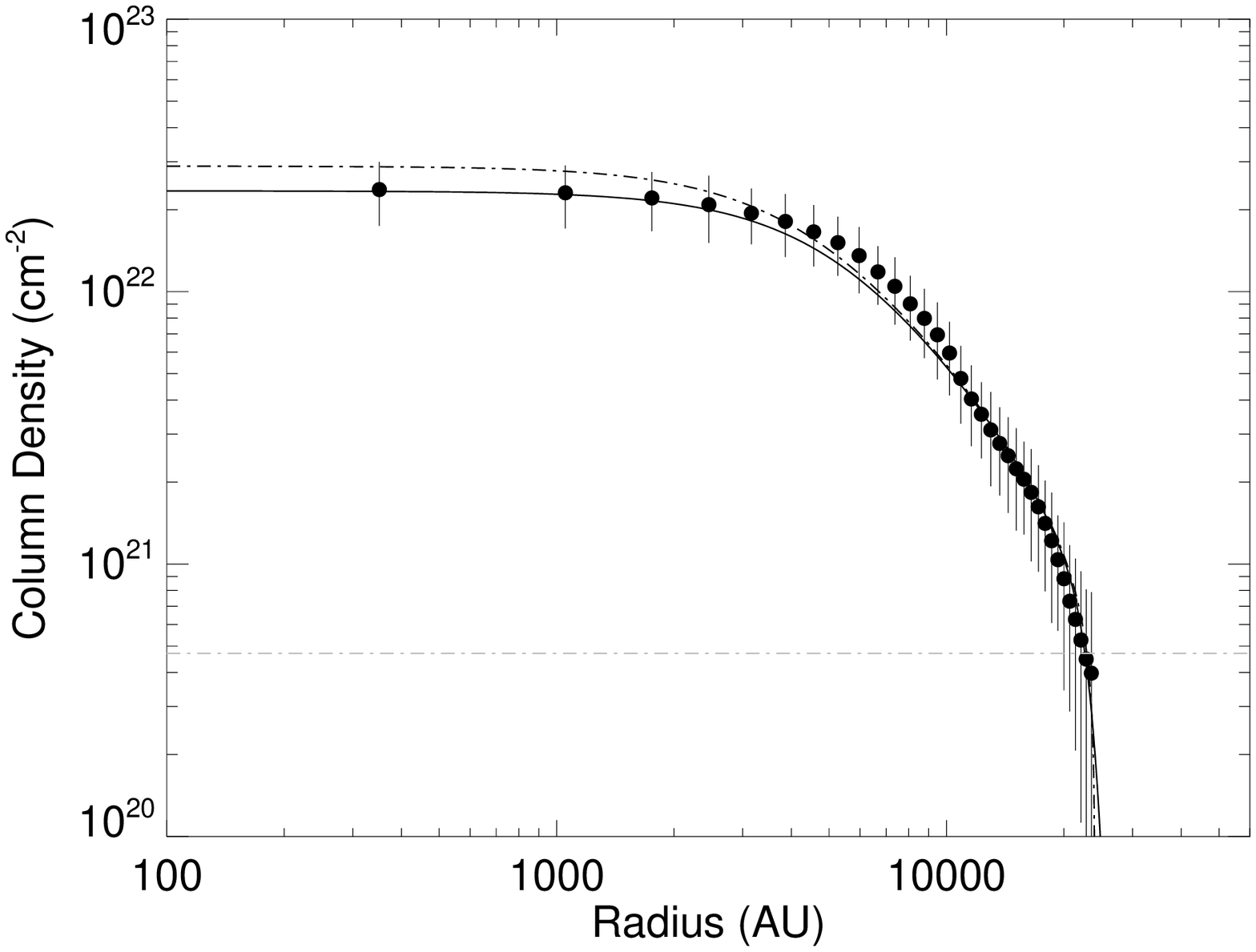}
\end{center}
 \caption{The radial column density profile and Bonnor--Ebert model fit for CB81. The dots represent the circularly averaged $N({\rm H}_2)$ values at each annulus arranged at 4$.\hspace{-3pt}''$5 intervals, and the error bars represent the rms dispersion of data points in each annulus. The solid line indicates the best-fit Bonnor--Ebert profile which is convolved with the same beam ($34''$ resolution) used in the $A_V$ measurements. The dashed line denotes the Bonnor--Ebert profile before the convolution. The gray dot-dashed lines denote the $1\sigma$ deviations of the column density in the reference field.}
   \label{fig5}
\end{figure}

\clearpage

\begin{figure}[t]  
\begin{center}
 \includegraphics[width=6.5 in]{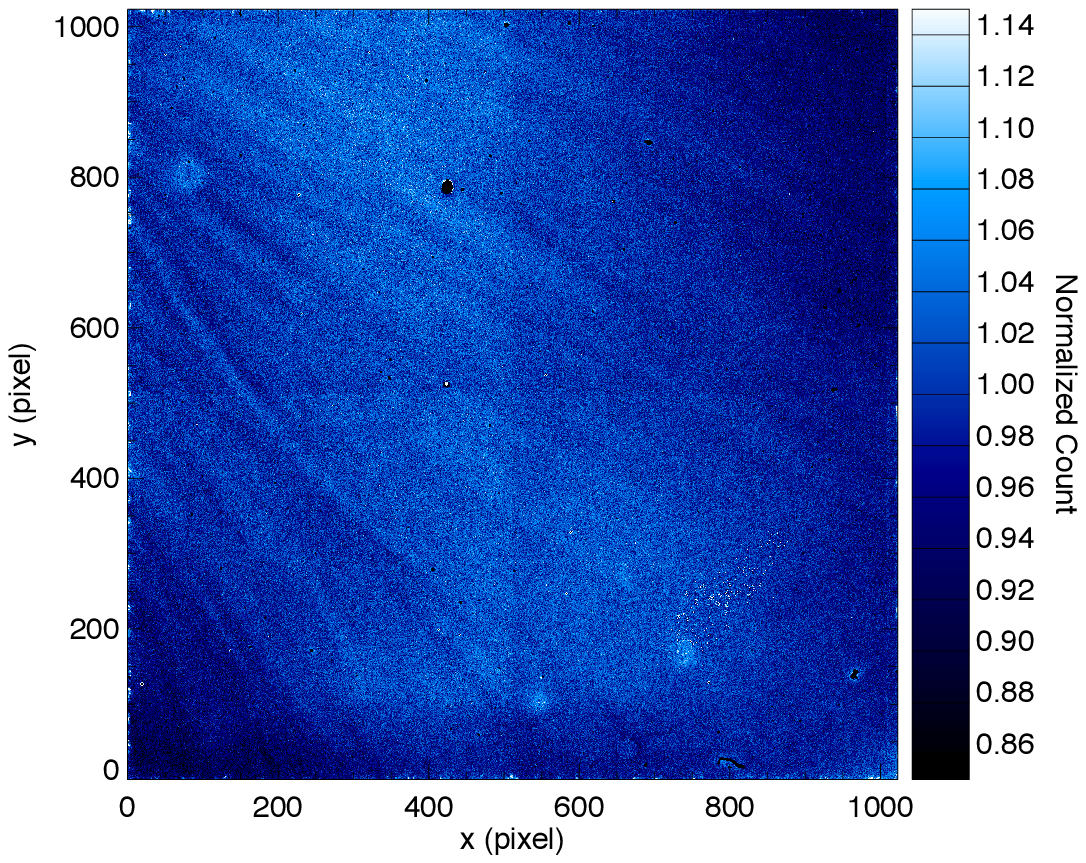}
\end{center}
 \caption{
The twilight flat image in the $H$ band. 
}
\end{figure}

\clearpage

\begin{figure}[t]  
\begin{center}
 \includegraphics[width=6.5 in]{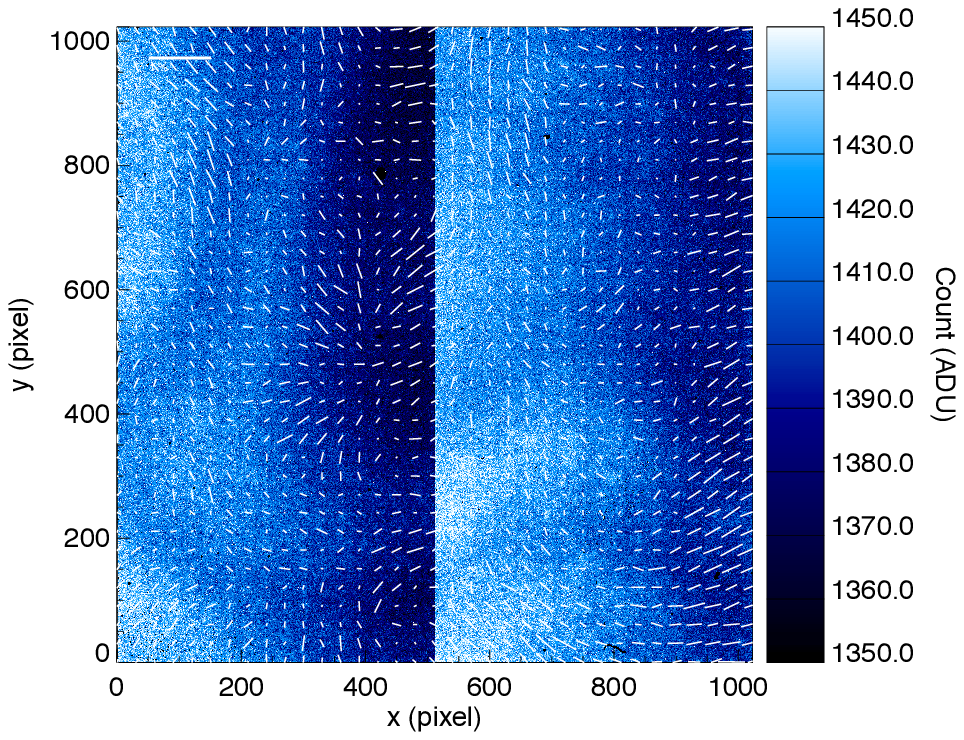}
\end{center}
 \caption{
The polarization vector distribution (white vectors) of the median sky image in the $H$ band. The background image is the Stokes I image of the median sky. The scale bar in the upper-left corner indicates 1\% polarization. 
}
\end{figure}

\clearpage

\begin{figure}[t]  
\begin{center}
 \includegraphics[width=6.5 in]{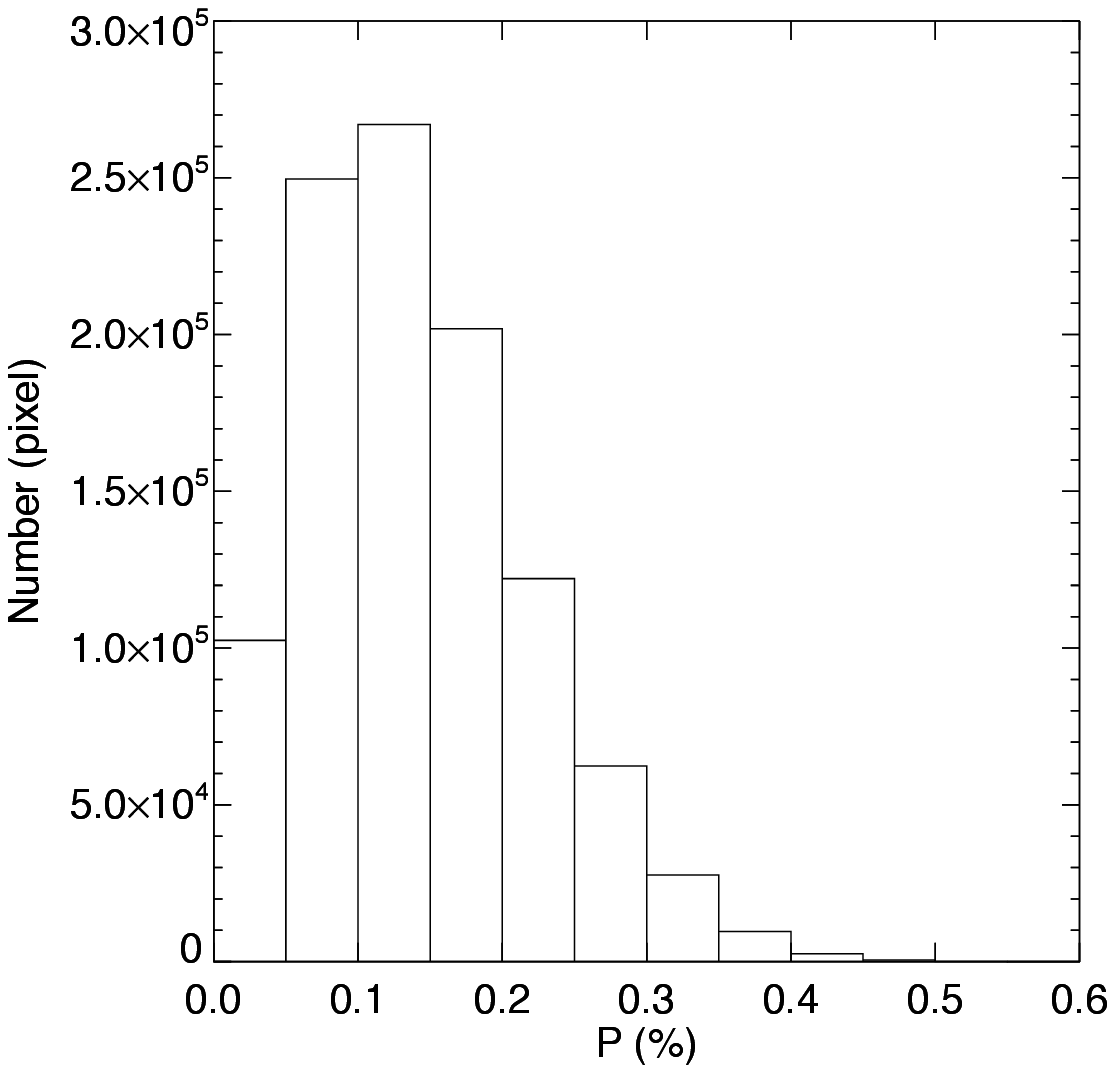}
\end{center}
 \caption{
A histogram of the $H$ band polarization image map shown in Figure S4.
}
\end{figure}

\end{document}